\documentclass[structabstract]{aa}

\usepackage{graphicx}
\usepackage{amssymb,amsmath}
\usepackage{natbib}
\usepackage{txfonts}

\def\Msun{ M_\odot}
\def\Rsun{ R_\odot}
\def\Mjup{ M_{\rm J}}

\def\Mp{M_{\rm p}}
\def\Rp{R_{\rm p}}
\def\Mi{M_{i}}
\def\Ri{R_{i}}
\def\Ms{M_{\star}}
\def\Rs{R_{\star}}

\def\Mearth{ M_\oplus}
\def\Rearth{ R_\oplus}
\def\Op{\Omega_{\rm p}}
\def\Os{\Omega_{\star}}

\def\Tp{T_{\rm p}}
\def\Ts{T_\star}

\def\G{\mathcal{G}}

\usepackage[normalem]{ulem}
\usepackage{color}
\definecolor{blue}{RGB}{0,0,255}
\definecolor{red}{RGB}{255,0,0}
\definecolor{green}{RGB}{0,200,0}
\definecolor{black}{RGB}{0,0,0}

\begin{document}

\title{Tidal dissipation and eccentricity pumping:\\ Implications for the depth of the secondary eclipse of 55~Cnc~e}

   \subtitle{ }

   \author{ Emeline Bolmont \inst{1,2}
	\and Franck Selsis \inst{1,2} 
          \and Sean N. Raymond \inst{1,2} 
          \and Jeremy Leconte \inst{3}
          \and Franck Hersant \inst{1,2}
          \and Anne-Sophie Maurin \inst{1,2} 
          \and Jessica Pericaud \inst{1}
                }

   \institute{Univ. Bordeaux, LAB, UMR 5804, F-33270, Floirac, France
   			 \and CNRS, LAB, UMR 5804, F-33270, Floirac, France
         \and Laboratoire de M\'et\'eorologie Dynamique, Institut Pierre Simon Laplace, Paris, France}

   \date{Received xxx ; accepted xxx}


  \abstract
{} 
{We use the super Earth 55~Cnc~e as a case study to address an observable effect of tidal heating. This transiting short-period planet belongs to a compact multiple system with massive planets. We investigate whether planet-planet interactions can force the eccentricity of this planet to a level affecting the eclipse depth observed with Spitzer.  }
{Using the constant time lag tidal model, we first calculate the observed planet flux as a function of albedo and eccentricity, for different tidal dissipation constants and for two extreme cases: a planet with no heat redistribution and a planet with full heat redistribution. We derive the values of albedo and eccentricity that match the observed transit depth. We then perform N-body simulations of the planetary system including tides and General Relativity to follow the evolution of the eccentricity of planet e. We compare the range of eccentricities given by the simulations with the eccentricities required to alter the eclipse depth.} 
{Using our nominal value for the dissipation constant and the most recent estimates of the orbital elements and masses of the 55~Cnc planets, we find that the eccentricity of planet e can be large enough to contribute at a measurable level to the thermal emission measured with Spitzer. This affects the constraints on the albedo of the planet, which can be as high as $0.9$ (instead of $0.55$ when ignoring tidal heating). We also derive a maximum value for the eccentricity of planet e directly from the eclipse depth: $e<0.015$ assuming Earth's dissipation constant.}
%
{Transiting exoplanets in multiple planet systems -- like 55 Cancri -- are exceptional targets for testing tidal models because their tidal luminosity may be observable. Future multi-wavelengths observations of eclipse depth and phase curves (for instance with EChO and JWST) should allow us to better resolve the temperature map of these planets and break the degeneracy between albedo and tidal heating that remains for single band observations. In addition, an accurate determination of the eccentricity will make it possible to constrain the dissipation rate of the planet and to probe its internal structure.}
		
\titlerunning{Tidal dissipation and eccentricity pumping}
\maketitle
%

\section{Introduction}

The bright star 55~Cancri hosts at least five planets \citep{Fischer2008} including the close-in $8M_{\oplus}$, $2.17R_{\oplus}$ planet e that is on a transiting orbit \citep{Demory2011,Winn2011,Gillon2012}. Table \ref{table1} shows the parameters of the star and planet e. Planets 55~Cnc~e and Kepler~10~b are the only low-mass planets ($M<10M_{\oplus}$) whose secondary eclipse has been observed, respectively by Demory et al. (\citeyear{Demory2012}, hereafter D12) and Batalha et al. \citeyearpar{Batalha2011}. In the case of Kepler~10~b, the planetary signal responsible for the eclipse depth is dominated by reflected light due to the spectral window of Kepler ($400-900$~nm). For this planet, a high albedo ($A>0.5$) is required to fit the observation \citep{Batalha2011,Rouan2011}. The secondary eclipse of 55~Cnc~e has been measured by D12 using the 4-5~$\mu$m IRAC2 channel of Spitzer. At these wavelengths, the thermal emission dominates the planetary flux. The observation implies an upper limit on the albedo ($A<0.55$) as well as an inefficient heat transport towards the night side of the planet. Even with a null albedo, a planet with a uniform temperature would have an insufficient dayside emission to produce the observed drop of flux during the secondary eclipse (D12). The required albedo depends of course strongly on the radius of the planet but Gillon et al. \citeyearpar{Gillon2012} have put tight constraints on the radius of 55~Cnc~e by combining primary transit data from both MOST and Spitzer. The parameters of the bright host star 55~Cnc have been accurately characterized using various techniques (including interferometric measurements) by \citet{VonBraun2011}. 

\begin{center}
\begin{table*}[bt]
\hfill{}
\begin{tabular}{ccccccc}
\hline
\hline
Star & $M$ ($\Msun$) & Radius  ($\Rsun$)  & P (days) & ${\rm log}(g)$ & $T_{{\rm eff}}$(K)&[Fe/H]\\
\hline
         & $0.905$     & $0.943$    & $39$ &  $4.45$& $5196$ & $0.31$\\
         &&&&&\\
\hline
\hline
Planet & $M$ sin $i$ & Radius & $a$ (AU)   &  & &\\
\hline
55~Cnc~e & $8.26~\Mearth$ & $2.17~\Rearth$ &  $0.01560$ &  & &\\
\hline
\end{tabular}
\hfill{}
\caption{Stellar properties and minimum mass, radius, and semi-major axis of planet 55~Cnc~e.
The stellar mass, radius, effective temperature, gravity and metallicity come from \citet{VonBraun2011} and its rotation period comes from \citet{Fischer2008}.
The values of $M$ sin $i$ and $a$ come from table 10 of \citet{Dawson2010}. The radius of 55~Cnc~e comes from \citet{Gillon2012}.}
\label{table1}
\end{table*}
\end{center}

Planets Kepler~10b and 55~Cnc~e share a lot of similarities but their surface albedo derived from secondary eclipse seem to differ. The composition of the two planets can of course be different. In particular, 55~Cnc~e has a surprisingly low density that implies a significant volatile content (and therefore a gaseous envelope), as suggested by D12, or a carbon-rich composition (ceramics being less dense than silicates) as suggested by \citet{Madhu2012}.  In the present study, we investigate a scenario that could change the albedo determination, in which the thermal emission of 55~Cnc~e is enhanced by the tidal dissipation resulting from a low eccentricity maintained by the other planets of the system. With a method similar to the one \citet{Barnes2010} used to infer the eccentricity of Corot-7 b and its tidal heating, we used N-body simulations that include tidal interactions and General Relativity to study the evolution of the system and of 55~Cnc~e in particular. We find that the orbit of 55~Cnc~e cannot be fully circularized and keeps a low eccentricity subjected to important variations. Using the planets' orbital elements derived from radial velocity data, we find that the eccentricity of planet e can reach values close to 0.01, which is below the upper limit from observations (e=0.06, D12) but enough for dissipation to produce an internal heat flux comparable to the stellar heating. In this case, the tidal dissipation affects the observed secondary eclipse depth, which can be compatible with a high albedo.\\
In a first section we study the influence of tidal dissipation on the thermal emission of the planet and derive the minimum eccentricities required to affect the thermal phase curve and secondary eclipse of 55~Cnc~e, for two extreme cases: a planet with no heat redistribution - no atmosphere - and a planet with a full redistribution of the stellar heating. In a second section, we simulate the dynamical evolution of the 55~Cnc system in order to constraint the range of possible eccentricities for planet e.


\section{Effect of eccentricity on the secondary transit depth.}
\label{section1}


\subsection{Modeling the planet flux}

In order to bracket the range of possible situations we consider two different cases: a synchronously-rotating planet with no atmosphere and a planet with a uniform temperature due to a fully efficient redistribution of the stellar heating by an atmosphere. \\
The bolometric stellar flux $\phi_{\star}$ and the spectral flux density $\phi_{\star,\lambda}$ received at a distance $d$  are given by given by
$$\phi_{\star}(d)=\sigma T_{eff}^4 \left(\frac{R_{\star}}{d}\right)^2,$$
where $\sigma$ is the Stefan-Boltzmann constant, $T_{eff}$ and $R_{\star}$ are the effective temperature and radius of the star, and
$$\phi_{\star,\lambda}(d)=F_{\star,\lambda}\left(\frac{R_{\star}}{d}\right)^2,$$
where $F_{\star,\lambda}$ is the spectral flux density at the surface of the star and is computed with the stellar atmosphere code Phoenix \citep{Allard2012} for the $T_{\textrm{eff}}$, gravity (log($g$)) and metallicity of 55~Cnc (see Table \ref{table1}). This synthetic spectrum agrees well with available observations (I. Crossfield, private communication), which are numerous for such a bright star (e.g. IRAS, IRTF, Akari, Spitzer, Wise).

The planet spectrum is a combination of its thermal emission and the reflected light. To derive the thermal component we first need to calculate the surface temperature map, which depends on both the absorbed stellar flux and the internal heat flux. Note that we consider an eccentricity that remains small enough so that its effect on the irradiation remains negligible. To calculate the stellar heating, we can thus assume a circular orbit and a synchronous rotation. In those circumstances, the irradiation at a given point of the planet does not vary with time. We checked with a more detailed thermophysical model that accounting for the small librations of the planet and the thermal inertia of the surface does not change our results. \\
If we assume no redistribution of the incident flux (no atmosphere), then the stellar flux absorbed by the planetary surface is a function of the zenith angle $\theta$ and the Bond albedo $A_B$ and is given by $$\phi_{\textrm{abs}}=\phi_{\star}(a)\cos\theta (1-A_B),$$
$a$ being the semi-major axis of the orbit. 
As we set the origin of the geographic coordinates at the substellar point, the zenith angle is given by $\cos\theta=\cos(\textrm{latitude})\cos(\textrm{longitude})$. We also include an internal heat flow due to tidal dissipation. Other sources of internal heat like radioactivity or remnant of accretion/differentiation can dominate when dissipation is negligible but in this case the resulting surface warming would be too low to affect the observed emission. For instance, the mean heat flow of the Earth corresponds to a surface temperature of only $30$~K. Significantly higher values would imply a very young planet, a case we do not consider here given \citet{VonBraun2011}'s estimation of the age of 55 Cnc: $10.2$~Gyr.  Locally, the flux emitted by the surface is equal to the sum of the absorbed stellar flux and the internal heat flow $\phi_{\textrm{tides}}$ resulting from tidal dissipation, which we assume uniform over the whole planetary surface. On the dayside, the surface temperature is thus given by
$$T_\textrm{day}(\theta)=\left(\frac{\phi_\textrm{abs}(\theta)+\phi_{\textrm{tides}}}{\sigma}\right)^\frac{1}{4},$$ while on the night side we have
$$T_{\textrm{night}}=(\phi_{\textrm{tides}}/\sigma)^\frac{1}{4}.$$
In the case of a full redistribution of the absorbed stellar energy the temperature is uniform over the whole planet and simply given by 
$$T_{\textrm{unif}}=(T_{\textrm{eq}}^4+\frac{\phi_{\textrm{tides}}}{\sigma})^\frac{1}{4},$$ where $T_{\textrm{eq}}$ is the equilibrium temperature of the planet:
$$T_{\textrm{eq}} = \left(\frac{\phi_{\star}(a) (1-A_B)}{4\sigma}\right)^\frac{1}{4}.$$
The surface of the planet is divided into a longitude-latitude grid. Each cell $j$ of the grid has a surface temperature $T_j$ and an area $S_j$. The flux spectral density received by a distant observer at a distance $d$ is given by 
$$ \phi_{P,\lambda}(d) = \sum_j I_j \frac{S_j\cos\alpha_j}{d^2},$$
where $\alpha_j$ is the angle between the normal to the cell and the direction toward the observer and $I_j$ is the specific intensity of the cell given by
$$I_j = \frac{\epsilon_{\lambda} B_{\lambda}(T_j) + A_\lambda \phi_{\star,\lambda}(a) \cos\theta_j}{\pi},$$
where $B_{\lambda}$ is the Planck function, $\epsilon_\lambda$ is the surface emissivity, and $A_\lambda$ is the surface spectral albedo. In this study, we use $\epsilon_\lambda=1$ and $A_\lambda= A_B=A$. In practice, $\cos\alpha_j$ is calculated as the dot product of the directions vectors attached to the center of the planet and pointing towards the observer and the center of the $j$ cell. Only locations visible to the observer ($\cos\alpha_j> 0$) contribute to $\phi_{P,\lambda}$.

\begin{figure}[ht]
	\centering
	\includegraphics[width=\linewidth]{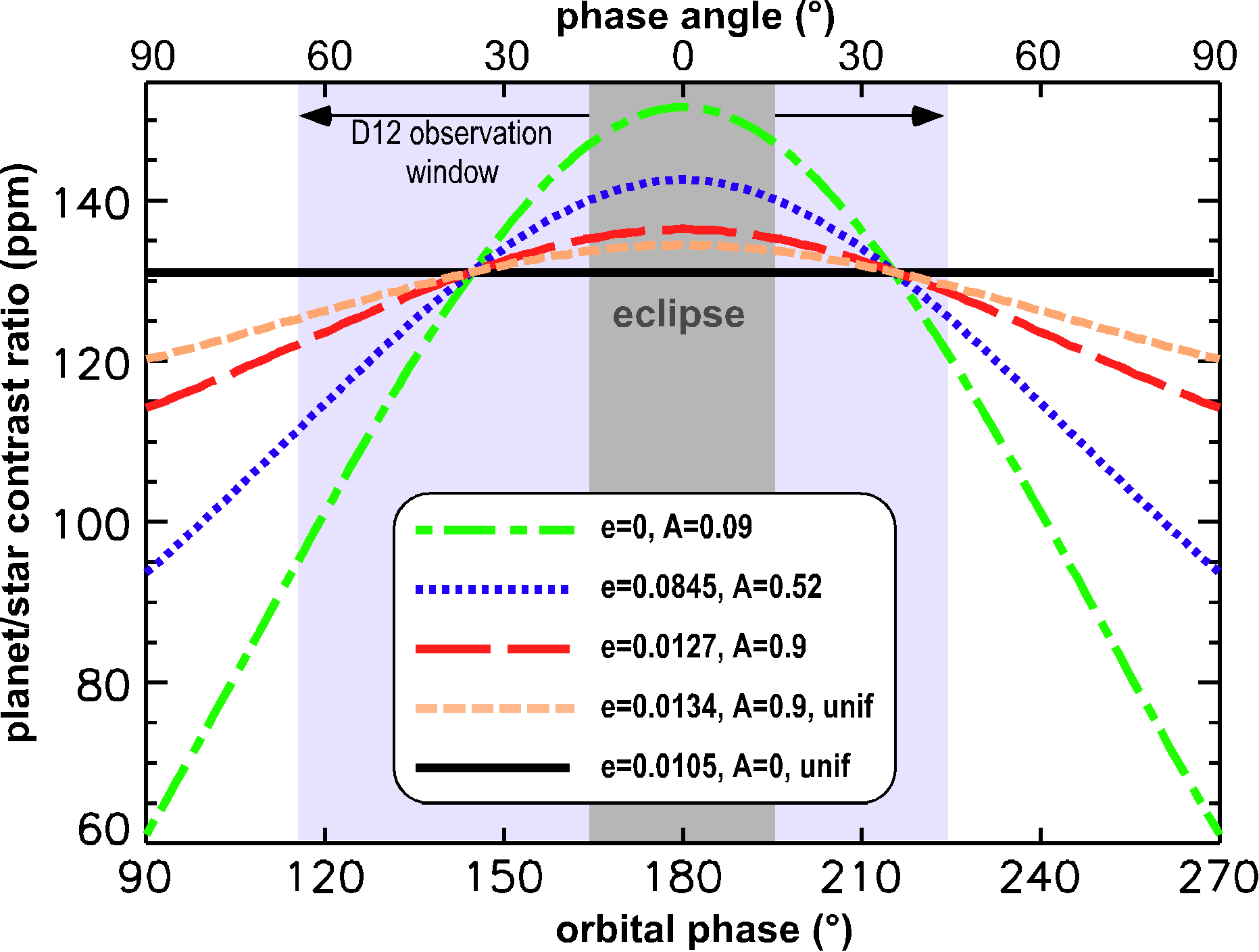}
 	\caption{The effect of albedo and eccentricity on the light curve. This graph shows the planet/star contrast ratio as a function of the orbital phase for different values of $(e,A)$. All these models give the same flux (131~ppm) for a phase angle of $35^\circ$, which is the average value for the out-of-eclipse Spitzer observations by D12. Calculations are done with the nominal value $\sigma_p$ and the orbital parameters and planetary mass from \citet{Endl2012}. The short-dashed and full lines are for a uniform temperature (perfect redistribution of heat) while there is no redistribution in the three other cases.}
	\label{fig:transit_sim}%
\vspace{5pt}
	\includegraphics[width=\linewidth]{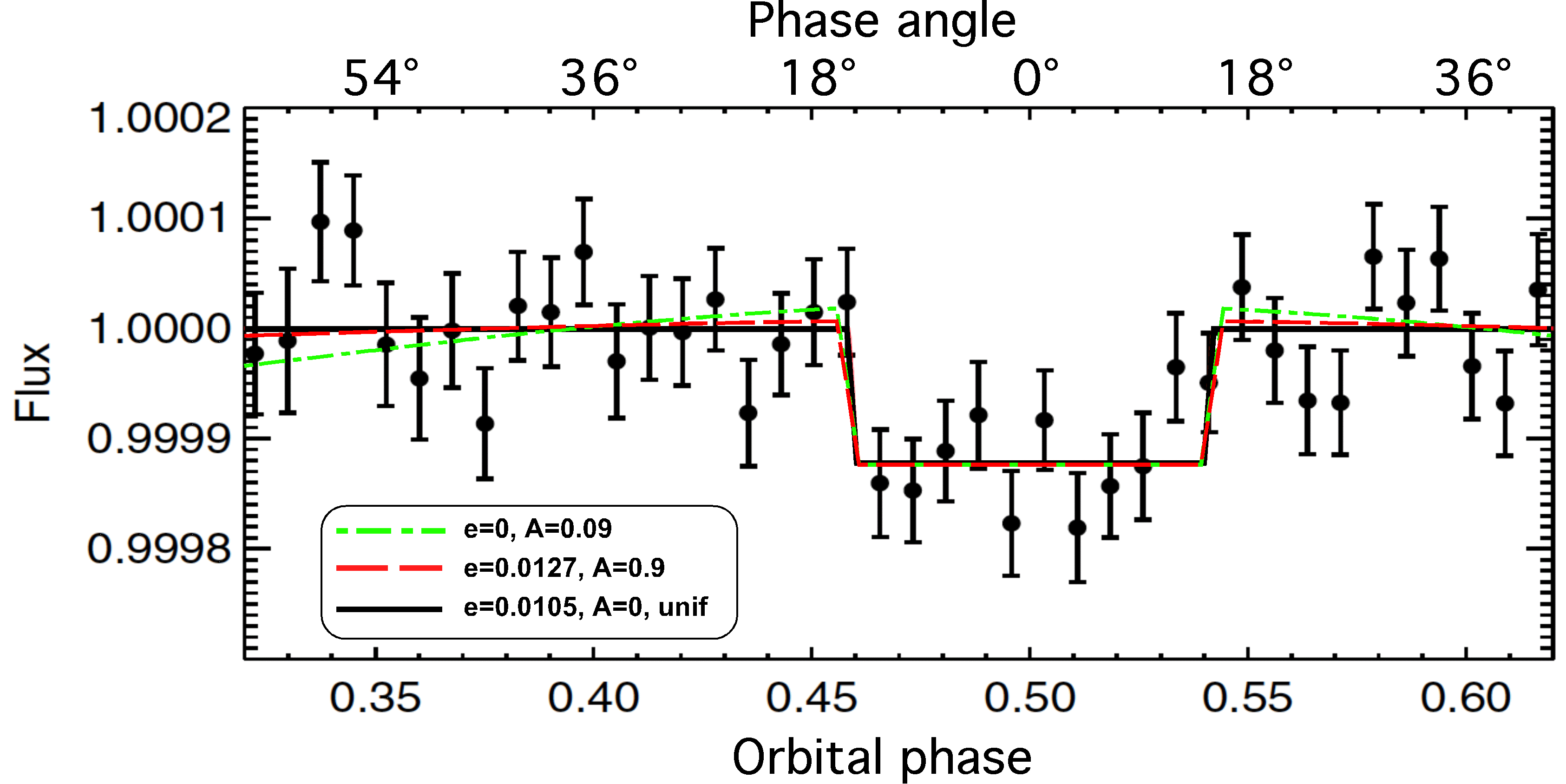}
 	\caption{Same as Fig.~\ref{fig:transit_sim} but the modeled light curves are superimposed to the actual photometric observations with Spitzer as published by D12. Only 3 of the cases from in Fig.~\ref{fig:transit_sim}, with the same line styles and color, are shown here.}
	\label{fig:transit_sim_D12}%
	\end{figure}

\begin{figure*}[ht]
	\centering
	\includegraphics[width=17cm]{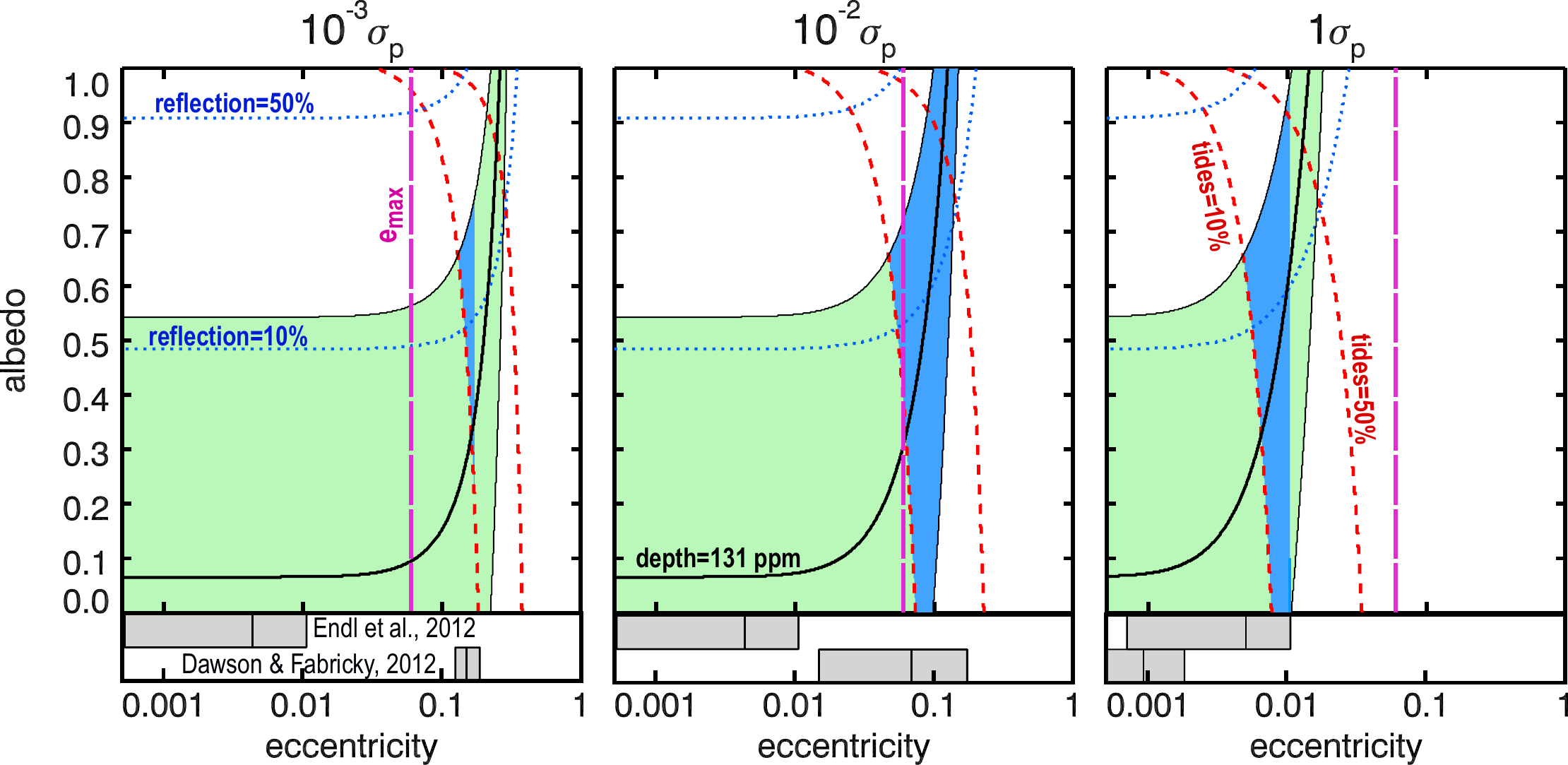}
 	\caption{Contraints on albedo and eccentricity from the secondary eclipse depth, in the case of an airless planets. The light green shaded areas are in agreement with the $131\pm28$~ppm depth measured with IRAC2/Spitzer by D12. The right hand side plot is for an Earth-like dissipation constant $\sigma_p$ ;  the left and middle plots are for $10^{-3}\times\sigma_p$ and $10^{-2}\times\sigma_p$, respectively. The dotted (resp. dashed) lines indicate where reflection (resp. tidal dissipation) accounts for 10 and 50\% of the planetary flux (reflection + thermal re-emission + tides = 100\%). The long dashed pink vertical line represents the observational constraint on the eccentricity given by D12. Horizontal grey bars below the graphs indicate the range of eccentricities found with the dynamical simulations described in section \ref{multi}. The upper and lower bars correspond to initial orbital elements from \citet{Endl2012} and \citet{Dawson2010}, respectively. The maximum and mean values of the eccentricity are indicated, as well as the minimum when larger than $5\times10^{-4}$. The blue areas are in agreement with the transit depth, the condition that tides accounts for more than $10\%$ of the planetary flux and compatible with an eccentricity within the range found with the dynamical simulations. For $10^{-3}\times\sigma_p$ and $10^{-2}\times\sigma_p$ the agreement is met for simulations of \citet{Dawson2010} while for $1\times\sigma_p$ the agreement is met for \citet{Endl2012}.}
	\label{fig:A_e}%
	\end{figure*}

\begin{figure*}[ht]
	\centering
	\includegraphics[width=17cm]{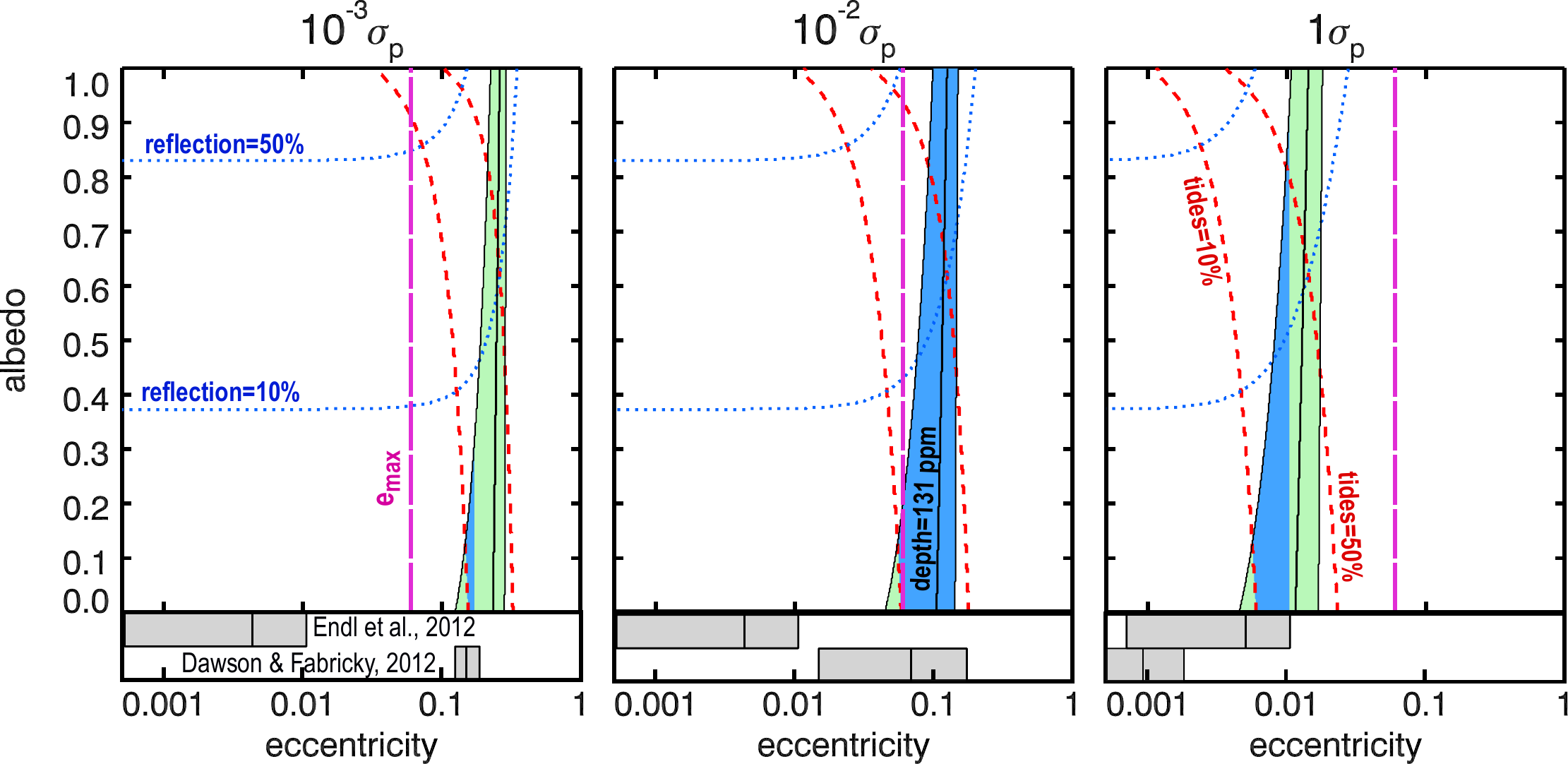}
 	\caption{Same as in Fig.~\ref{fig:A_e} but for an isothermal planetary surface.}
	\label{fig:A_e_unif}%
	\end{figure*}

In the previous formula, the star is considered as a point source but this approximation can become inadequate in the case of extremely close planets. Seen from 55~Cnc~e, the angular radius of the stellar disk is $15^{\circ}$. This has two main consequences. First, it affects the distribution of the stellar flux over the planetary surface and thus the temperature map. The terminator that delimits the dayside and the nightside is replaced by a penumbra region, as described by L\'{e}ger et al. \citeyearpar{Leger2011}. This effect, however, has a negligible impact on the observed flux even in the case of 55~Cnc~e because the affected regions of the planet emit/reflect an insignificant fraction of the total spatially-unresolved planetary flux. We simulated numerically an extended star by spreading the stellar luminosity over a number of fainter punctual sources distributed over the stellar disk. We found that, for phase angles smaller than $120^{\circ}$, the planetary flux calculated with a point source star and with an extended star differ by less than 1\%. There is a second, more significant, effect: phase angles smaller than $15^{\circ}$ cannot be observed as they are eclipsed by the star. The out-of-transit photometric reference used by D12 actually covers a broad range of phase angles: $68-15^{\circ}$ before the eclipse and $15-43^{\circ}$ after, while the eclipse itself lasts about one tenth of the orbit. With our model, we can calculate the planetary flux from an airless planet for any phase angle (i.e. sub-observer longitude). Therefore, we can either produce the light curve for the whole range of observed phase angles (as in Fig.~\ref{fig:transit_sim} to \ref{fig:transit_sim_D12}) or for a phase angle of  $35^{\circ}$, which is the averaged value of the out-of-transit observations by D12 (as in Fig.\ref{fig:A_e} to \ref{fig:A_e_unif}). 

To compare the computed planetary flux with the measured transit depth of $131 \pm 28$~ppm derived by D12, we calculate the planet to star contrast ratio $R$ as follows:
$$R=\frac{\int_{\Delta\lambda} \phi_{P,\lambda}(D) w_\lambda \, \mathrm{d}\lambda}{\int_{\Delta\lambda} \phi_{\star,\lambda}(D) w_\lambda \, \mathrm{d}\lambda},$$
where $D=12.34$~pc is the distance of the 55~Cnc system, $w_\lambda$ is the spectral response of the IRAC2 detector (as given in Fig. 2.4 of the IRAC Instrument Handbook), and $\Delta \lambda$ is the IRAC2 band ($3.9-5.1~\mu$m). 

\subsubsection{Tidal heating}
\label{tidemodel1}

The average surface heat flux $\phi_{\textrm{tides}}$  is given by $\dot{E}_{tides}/4\pi R_{P}^2$ where $\dot{E}_{tides}$ is the rate of tidal dissipation. We use here the constant time lag model to compute $\dot{E}_{tides}$. For a planet of mass $\Mp$ in pseudo-synchronous rotation orbiting a star of mass $\Ms$ at a semi-major axis $a$ and with a null obliquity \citep[formula 13 from][]{Leconte2010}, the rate of tidal dissipation is given by:
\begin{equation}
\dot{E}_{tide} = 2 \frac{1}{\Tp}\frac{\G \Mp \Ms}{4a} \left[Na1(e) - \frac{Na2(e)^2}{\Omega(e)} \right ],
\end{equation}
where the dissipation timescale $\Tp$ is defined as
\begin{equation}
\label{Tp}
\Tp = \frac{1}{9}\frac{\Mp}{\Ms(\Mp+\Ms)}\frac{a^8}{\Rp^{10}}\frac{1}{\sigma_{p}}
\end{equation}
and depends on the mass of the planet $\Mp$, its dissipation factor $\sigma_{p}$ (as defined by \citet{Hansen2010}) and of the mass of the star $\Ms$. $Na1(e)$, $Na2(e)$ and $\Omega(e)$ are eccentricity-dependent factors:

\begin{align*}
Na1(e) &= \frac{1+31/2e^2+255/8e^4+185/16e^6+85/64e^8}{(1-e^{2})^{15/2}},\\
Na2(e) &= \frac{1+15/2e^2+45/8e^4+5/16e^6}{(1-e^{2})^{6}},\\
\Omega(e) &= \frac{1+3e^2+3/8e^4}{(1-e^{2})^{9/2}}.
\end{align*}

The difference in the rate of tidal dissipation assuming synchronization or pseudo-synchronization for planet 55~Cnc~e is $3.5\%$ for $e=0.06$ (the upper limit of eccentricity from D12) and only $0.1\%$ for $e=0.01$.

The characteristics of planet e can be seen in Table \ref{table1}.
We used for this study the latest estimation for the radius of 55~Cnc~e from \citet{Gillon2012}. Both \citet{Dawson2010} and \citet{Endl2012} give comparable masses for 55~Cnc~e. 

As in \citet{Bolmont2011,Bolmont2012} in order to be consistent with the Earth dissipation inferred from D245 astrometric data \citep{DeSurgyLaskar1997}, our nominal value of $\sigma_{p}$ is $1.877 \times 10^{-51}$~g$^{-1}$cm$^{-2}$s$^{-1}$. This value will be referred thereafter as 55~Cnc~e nominal dissipation factor.

The uncertainty on the dissipation factor is, however, considerable. The composition of the planet and its internal structure are unknown while they affect, and are affected by, the dissipation. For instance, if we assume that a melted planet dissipates less than a solid one, an initially solid planet can start to melt due to tidal heating, which in turn lowers its dissipation and the internal heating.   For a given eccentricity, this feedback may result in a steady state, in which the internal structure and mineral phases are consistent with the tidal heating. A self-consistent treatment would thus include a feedback between the dissipation rate and the dissipation factor, but that is beyond the scope of this study. (Such modeling has been done by \citet{Henning2009} and \citet{Behounkova2011}.)  Nonetheless, there is no reason that 55~Cnc~e's dissipation factor should be the same as Earth's. To take into account this uncertainty we performed simulations over 7 decades of dissipation factors from $10^{-5}$ to $10^2\times\sigma_{p}$.

	\begin{figure}[h!]
	\begin{center}
	\includegraphics[width=9cm]{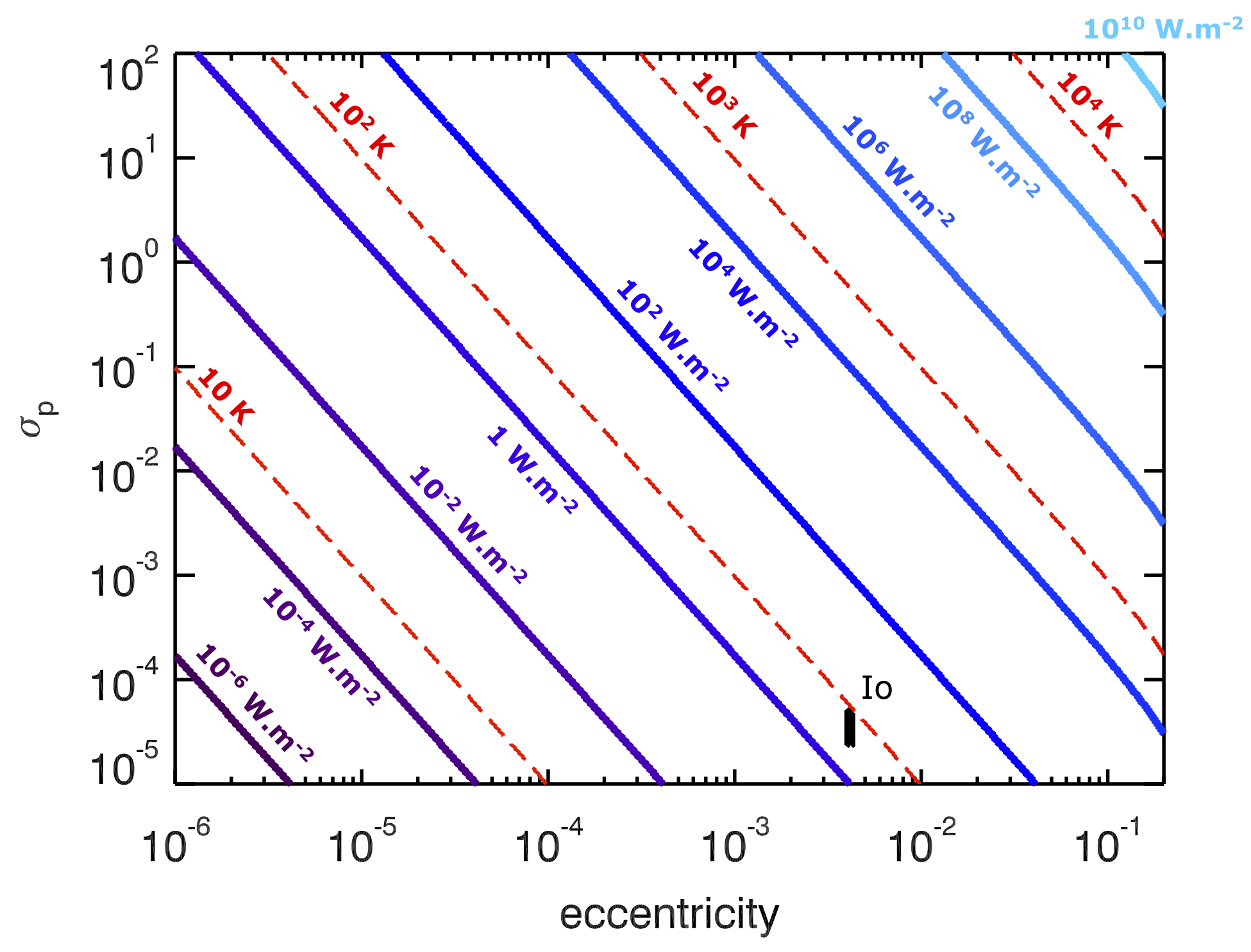}
	\caption{Map representing contours of the average surface heat flux $\phi_{\textrm{tides}}$, as a function of eccentricity and planetary dissipation factor. From left to right the full lines correspond to a flux of $10^{-6}$, $10^{-4}$, $10^{-2}$ $1$, $10^2$, $10^4$, $10^6$, $10^8$, $10^{10}$~W.m$^{-2}$. The long dashed lines correspond to a temperature $T=(\phi_{\textrm{tides}}/\sigma)^\frac{1}{4}$ of $10$, $10^2$, $10^3$ and $10^4$~K. The observed value of $\phi_{\textrm{tides}}$ on Io, $2.4-4.8$~W/m$^2$ \citep{Spencer2000}, is indicated in black for its estimated eccentricity value of $4.3\times 10^{-3}$ \citep{MurrayDermott}.}
	\label{map_tidal_flux}
	\end{center}
	\end{figure}

Figure \ref{map_tidal_flux} shows the contours of the average surface heat flux $\phi_{\textrm{tides}}$, as a function of eccentricity and dissipation factor. To affect the observation, $\phi_{\textrm{tides}}$ must be extremely high and comparable to the insolation flux. Such values exceed by orders of magnitude the tidal heat flux measured at the surface of Io, which is the largest in the Solar System. The highest values on Fig.~\ref{map_tidal_flux} certainly imply a melted surface and interior, as well as an extremely efficient vertical heat transfer, which may not be consistent with the assumed strong dissipation, as discussed above. Note, however, that a melted surface does not contradict the airless assumption as melted refractory materials have very low vapor pressures: \citet{Leger2011} calculated that the equilibrium pressure above a substellar magma ocean on Corot~7~b would not exceed a few Pa, with a negligible effect on the opacity and horizontal heat transport.


\subsection{Model vs observations}
\label{results1}

Besides parameters that are constrained by observations other than secondary eclipse ($M_\star, M_P, R_P, a$), $\phi_{\textrm{tides}}$ depends on the eccentricity and on the internal dissipation constant $\sigma_p$, while the albedo $A$ determines the ratio between reflected and absorbed stellar fluxes. At low eccentricities yielding negligible dissipation, the planetary flux depends only on the albedo (and on the heat redistribution assumption). For higher eccentricities, the resulting dissipation can increase the temperature and the flux depends on two additional parameters: $e$ and $\sigma_p$.\\
Figures~\ref{fig:A_e} and \ref{fig:A_e_unif} show contours of the eclipse depth in the IRAC2/Spitzer band as a function of eccentricity and albedo, and for three values of the dissipation constant: $10^{-3}\times\sigma_p$, $10^{-2}\times\sigma_p$ and $1\times \sigma_p$. Figure~\ref{fig:A_e} is for a planet with no atmosphere while Fig.~\ref{fig:A_e_unif} is for an isothermal photosphere (full redistribution of the stellar heating). The shaded area gives the $A - e$ domain compatible with the observation by D12 ($131 \pm 28$~ppm). The eclipse depth is calculated for a phase angle of $35^\circ$ corresponding to the average phase angle during the out-of-eclipse observations by D12. Isothermal cases cannot match the observation even for $A=0$ (as noted by D12) because the dayside would be too cold and would produce an eclipse that is too shallow unless there is an internal source of heat. This isothermal case may not, however, be realistic for such a close-in planet, even in the presence of a thick atmosphere. Indeed, the radiative cooling timescale, which varies as $T^{-3}$, is likely to be shorter than the dynamical one due to the high equilibrium temperature of this planet ($\sim 1600$~K, for an albedo of 0.5). Even a moderate day-night temperature contrasts would make the dayside emission significantly stronger than the nightside one. Note also that a phase modulation exists even in the isothermal case due to the reflected light (unless $A=0$).  \\

The cases with no heat redistribution match the observation with either an albedo lower than 0.55 and no eccentricity (in agreement with D12) or with a higher albedo (up to $A=1$) and an eccentric orbit. When increasing (resp. decreasing) the dissipation constant, lower (resp. higher) values of the eccentricity are required to modify the eclipse depth by a same factor. 
In Fig. \ref{fig:A_e} and \ref{fig:A_e_unif} the blue shaded area corresponds to cases for which the following three conditions are met: the resulting transit depth is within the observational range, the tides account for more than $10\%$ of the planetary flux and the eccentricity is within the range of eccentricities found by the dynamical simulations of the system. For dissipations of $10^{-3}\times\sigma_p$ and $10^{-2}\times\sigma_p$, the range of eccentricity comes from a simulation of the system of \citet{Dawson2010} and for a dissipation of $1\times\sigma_p$, it comes from a simulation of the system of \citet{Endl2012}. For example, assuming that tides represent more than $10\%$ of the planetary flux, the observations would be in agreement with a planet of albedo between $0.25$ and $0.75$ assuming a dissipation of $10^{-3}\times\sigma_p$. As we do not know the dissipation of the planet nor its precise eccentricity the observations would not allow to determine the albedo of the planet. However, D12 gave an observational constraint on the eccentricity of 55~Cnc~e, it should not be higher than $0.06$. It allows to rule out some of the solutions found. For a dissipation of $10^{-3}\times\sigma_p$, there would no longer be agreement with \citet{Dawson2010} because the eccentricity is too high. For a dissipation of $10^{-2}\times\sigma_p$, there would be an agreement with \citet{Dawson2010} for a planet of albedo between $0.35$ and $0.7$. For a dissipation of $1\times\sigma_p$, there would still be an agreement with \citet{Endl2012} for a planet of albedo lower than $0.95$.

We define here the critical eccentricity as the eccentricity above which tides affect the transit depth by more than $10\%$ and for which the transit depth is of $103$~ppm. This critical eccentricity depends on the dissipation and is given in Table~\ref{table4} for the range of dissipation factors considered here. In Figures \ref{fig:A_e} and \ref{fig:A_e_unif}, the critical eccentricity corresponds to the eccentricity of the intersection of the dashed line corresponding to ``tides = $10\%$'' and the upper line of the shaded area. The maximum eclipse depth that we find without dissipation is 134~ppm with no atmosphere and 98~ppm in the isothermal case (to be compared with the $131\pm 28$~ppm found  by D12).  \\

Whatever the redistribution efficiency, the albedo, and for a dissipation constant of $1\times\sigma_p$, eccentricities higher than $0.015$ produce too large a planet flux and can be ruled out. However, D12 constrained the eccentricity of 55~Cnc~e to be less than $0.06$, which allows us to rule out some of the solutions we found. \\

Fig.~\ref{fig:A_e} and \ref{fig:A_e_unif} also show the range of eccentricities found with N-body simulations that are described in the next section. These N-body simulations use two different sets of orbital elements and masses derived from radial velocities, one from \citet{Dawson2010} and the other from \citet{Endl2012}. In particular, the value for the orbital distance of planet e slightly differs between these two sets (see Table\ref{table3}). The effect on the transit depth calculation is, however, very small and would hardly be seen on these plots. Therefore, Fig.~\ref{fig:A_e} and \ref{fig:A_e_unif} only present the results obtained with $a=0.01544$~AU, the value from \citet{Endl2012}. \\

As shown in Fig.~ \ref{fig:transit_sim}, orbital photometry (i.e. phase curve measurement) could break the degeneracy between the different sets of albedo and eccentricity. Cases with strong tidal heating and high albedo exhibit much flatter phase curves than cases with negligible tidal heating and low albedo. Distinguishing between these cases at 4.5~$\mu$m would require a photometric precision of the order of 10~ppm, which is not achieved yet as seen in Fig.~\ref{fig:transit_sim_D12}. The space telescope EChO \citep{EChO2012} should reach the required precision and could also observe at wavelengths as long as 11~$\mu$m and possibly 15~$\mu$m. At these wavelengths the planet/star contrast ratio would increase significantly (200~ppm at 11~$\mu$m). Also by measuring the eclipse depth and/or the phase curve at different wavelengths simultaneously, EChO could get a much better constraint on the temperature map of the planet and thus on the albedo vs eccentricity.

With no other constraints on the eccentricity, our results show that all values of the albedos between 0 and 1 are possible for 55~Cnc~e if its dissipation factor is of $10^{-2}\times\sigma_p$. The goal of the next section is to estimate the maximum eccentricity of planet e using N-body simulations of the 55~Cnc system.


\section{The eccentricity of 55~Cnc~e: damping vs forcing}

In a system with one planet, the eccentricity of such a close-in planet would be damped to zero very quickly, however 55~Cnc~e is part of a compact five-planet system. Planet-planet interactions can maintain an eccentricity, which would be an equilibrium between the tidal damping and the excitation by the other planets. 


\subsection{Eccentricity evolution assuming a single planet system}



The model used for the study of the evolution of one planet around its host star is a a re-derivation of the equilibrium tide model of \citet{Hut1981} as in \citet{EKH1998}. We consider both the tide raised by the star on the planet and by the planet on the star. We apply the constant time lag model \citep{Leconte2010}  and use Hansen's \citeyearpar{Hansen2010} estimation of the internal dissipation constant of giant exoplanets and their host stars for planet 'b' and 55~Cnc. For consistency, we use for planet e the same dissipation constant as in the tidal heating model (see section~\ref{tidemodel1}).

The secular evolution of the eccentricity of the planet is given by \citet{Hansen2010}: 

\begin{equation}\label{Hansene}
\begin{split}
\frac{\dot{e}}{e} &= \frac{1}{e}\frac{\mathrm{d}e}{\mathrm{d}t}\Big|_{plan} + \frac{1}{e}\frac{\mathrm{d}e}{\mathrm{d}t}\Big|_{star}\\
 &= -\frac{9}{2\Tp}\Big[Ne1(e)-\frac{11}{18}\frac{\Op}{n}Ne2(e)\Big]\\ 
& \quad - \frac{9}{2\Ts}\Big[Ne1(e)-\frac{11}{18}\frac{\Os}{n}Ne2(e)\Big],
\end{split}
\end{equation}
where $\Op$ is the planet rotation frequency, and $n$ is the mean orbital angular frequency. The stellar parameters are obtained by switching the $p$ and $\star$ indices. $Ne1(e)$ and $Ne2(e)$ are given by: 
\begin{align*}
Ne1(e) &= \frac{1+15/4e^2+15/8e^4+5/64e^6}{(1-e^{2})^{13/2}},\\
Ne2(e) &= \frac{1+3/2e^2+1/8e^4}{(1-e^{2})^{5}}.\\
\end{align*}

The integration of the tidal equations was performed using a fourth order Runge-Kutta integrator with an adaptive timestep routine \citep{Press1992}. The precision of the calculations was chosen such that the final semi-major axis of each integrated system was robust to numerical error at a level of at most one part in $10^3$.

The characteristics of the star and the inner planet of the 55~Cnc system are listed in Table \ref{table1}. 

We use for the star 55~Cnc the value of stellar dissipation derived by \citet{Hansen2010}: $\sigma_\star = 4.992 \times 10^{-66}$~g$^{-1}$cm$^{-2}$s$^{-1}$. We compute the tidal evolution of 55~Cnc~e with the dissipation factor given in section \ref{tidemodel1} and we also compute separately the tidal evolution of 55~Cnc~b. This planet is thought to be a gas giant, so Hansen's value for gas giants is used: $\sigma_b = 2.006 \times 10^{-60}$~g$^{-1}$cm$^{-2}$s$^{-1}$. The radius of 55~Cnc~b is calculated assuming a Jupiter density.

Because the planetary spin synchronization timescale is short compared to the other timescales considered here \citep{Leconte2010,HellerLeconteBarnes2011}, each planet rotation period is fixed to the pseudo-synchronization value at every calculation timestep. Note that for the low eccentricities found for planet e, the difference in the dissipation rates calculated assuming pseudo-synchronization and sychronization is negligible. If the obliquity of planet e is determined by planetary tides, it reaches zero on very short timescales. The obliquity of planet 'b' should not evolve tidally on less than Gyr timescales. For simplicity, we assume both obliquities to be zero.

\begin{figure}[h!]
	\begin{center}
	\includegraphics[width=9cm]{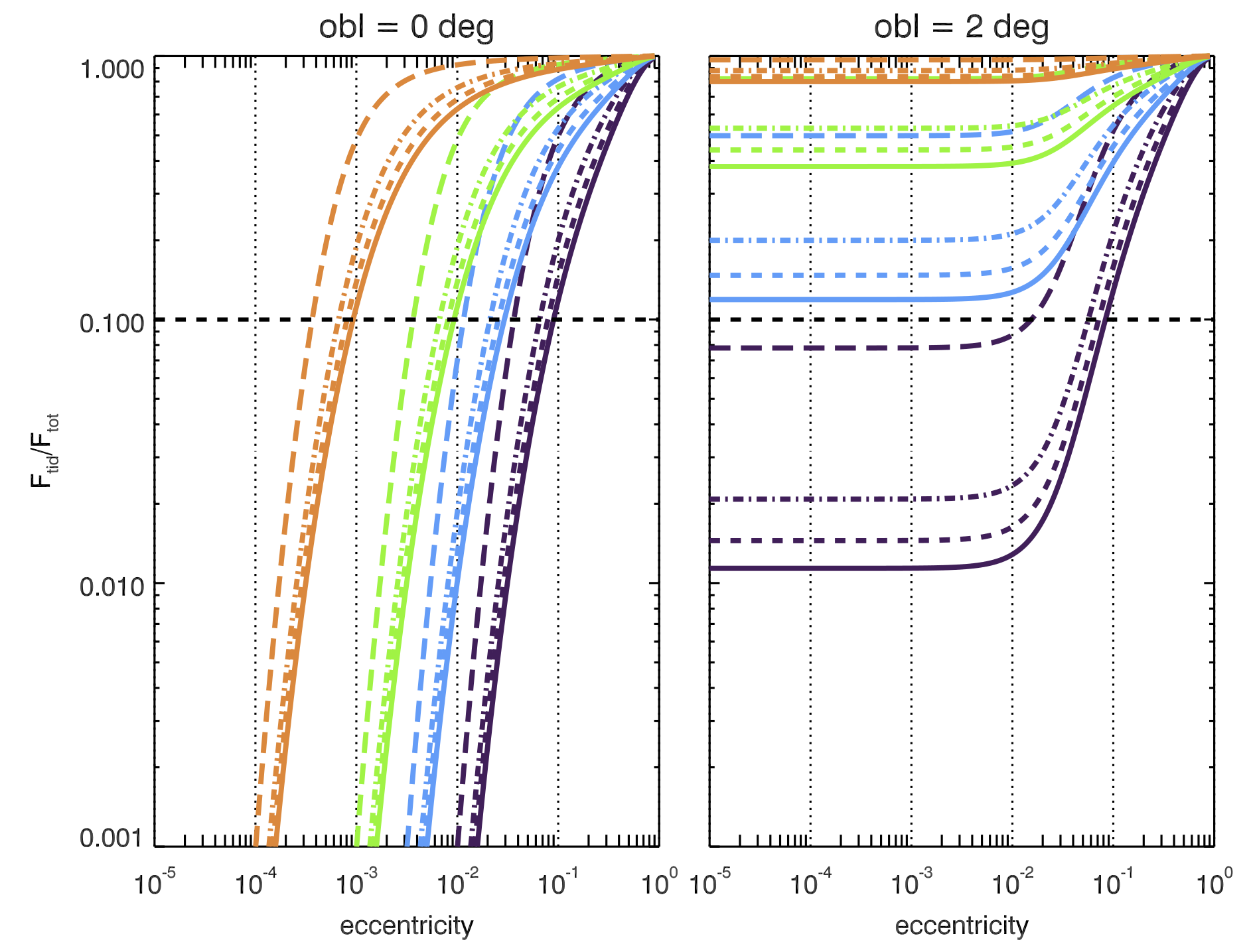}
	\caption{Ratio of 55~Cnc~e tidal flux over its total flux as a function of eccentricity for two different obliquities ($0^\circ$ and $2^\circ$). The purple curve corresponds to a dissipation factor of $10^{-3}\times \sigma_p$, the blue one to $10^{-2}\times \sigma_p$, the green one to $10^{-1} \times\sigma_p$ and the orange one to $10^{1} \times\sigma_p$. The full line corresponds to a Bond albedo of the planet of $0$, the small dashed line to an albedo of $0.5$, the dashed dotted line to an albedo of $0.8$ and the long dashed one to an albedo of $1$. The black horizontal dashed line represents a ratio of $10\%$.}
	\label{FtidoverFtot_0_2_spitz_real}
	\end{center}
	\end{figure}

If the obliquity of 55~Cnc~e were non-zero it would affect our results by increasing the planet's dissipation rate and tidal heat flux. Neglecting the obliquity of 55~Cnc~e therefore gives us a lower limit on the possible tidal flux. We calculated the tidal flux for two different obliquities of planet e. Fig. \ref{FtidoverFtot_0_2_spitz_real} shows the ratio of the planetary tidal- to total flux in the Spitzer band for different obliquities, for dissipation factors and albedos. As expected, the lower the dissipation factor, the higher the eccentricity needed to reach the ratio of $10\%$ of the planet tidal flux over its total flux. For a dissipation of $10^{-3}\times\sigma_p$, the eccentricity needed to reach $10\%$ is of about $0.1$ for a planet with an albedo of $0$ and this eccentricity is of $0.04$ if the albedo is of $1$. In the IRAC2 Spitzer band, the thermal emission dominates over the reflected light so increasing the albedo has an effect of decreasing the non tidal part of the flux emitted by the planet. Thus, a smaller eccentricity is needed for the tidal flux to reach $10\%$ of the planet total flux.

The right panel of Fig. \ref{FtidoverFtot_0_2_spitz_real} shows a case for which planet e's obliquity of $2$~degrees may have been excited by forcing from the other planets. Had this planet been alone, tides would have reduced this obliquity in very short timescales. The major difference with the left panel is that for small eccentricities, the ratio $F_{{\rm p,tides}}/F_{{\rm p,tot}}$ is non zero. The obliquity contributes to the tidal dissipation in a strong way. If the planetary dissipation is more than $10^{-2}\times\sigma_p$, the ratio is always bigger than $10\%$ whatever the value of the planet's albedo. For eccentricities bigger than $0.1$, the tidal dissipation due to the eccentricity becomes dominant again.

The study of the effect of obliquities of the inner planets on the N-body evolution of the system will be conducted in the future.

\label{damping}

The tidal evolution of a single planet around a star is determined by the tide raised by the planet on the star - the stellar tide - and the tide raised by the star on the planet -the planetary tide. 
The eccentricity damping timescale $\tau_{ecc}\big|_{plan}$ due to the planetary tide is given by $\tau_{ecc}\big|_{plan}   = \left|\frac{e}{\dot{e}}\right|_{plan}$.

For 55~Cnc~e, $\tau_{ecc}\big|_{plan} \sim 4600$~yrs, and the eccentricity damping timescale due to the stellar tide is $\tau_{ecc}\big|_{star} \sim 1.4\times10^{10}$~yrs. The evolution of the eccentricity of 55~Cnc~e will therefore be dominated by the planetary tide. Let us define the eccentricity damping timescale as $\tau_{ecc} = \tau_{ecc}\big|_{plan}$. In Fig. \ref{17191_evolution}, the evolutionary tracks of 55~Cnc~e are represented in purple dashed-dotted lines for a nominal dissipation factor $\sigma_p$. If this planet was alone in the system, the eccentricity would be damped in about $10 \times \tau_{ecc}\big|_{plan}$. 

For 55~Cnc~b, $\tau_{ecc}\big|_{plan}$ and $\tau_{ecc}\big|_{star}$ are longer than the age of the system. The eccentricity of 55~Cnc~b will remain at its present values on more than Gyrs timescales. In Fig. \ref{17191_evolution}, the evolutionary tracks of 55~Cnc~b are represented in blue dashed-dotted lines. If this planet was alone in the system, its semi-major axis and eccentricity would remain the same for more than one billion years.

However, to accurately describe the evolution of 55~Cnc~e, it is crucial to take into account the fact that these planets are interacting and are part of a five-planet system. 


\subsection{Eccentricity evolution assuming the five-planet system}
\label{multi}

\subsubsection{Tides in multiple planets systems}

To compute the tidal interactions we used the tidal force as expressed in \citet{Hut1981} for the constant time lag model. We added the force in the N-body code Mercury \citep{Chambers1999}. \citet{Kaib2011} showed that the system is expected to be nearly coplanar (all the orbits remain within $5^\circ$ in their simulations). \citet{Ehrenreich2012} claim that they have observed the signature of an extended hydrogen atmosphere of planet b, grazing the stellar disk, implying a very low mutual inclination between planets e and 'b'. Because planet e is transiting,  assuming coplanarity implies that all the minimum masses inferred from radial velocity are  close to the true masses. \citet{Kaib2011} predicted that the plane of the orbits should be misaligned with the star due to the influence of the binary companion 55~Cnc~B. In this study, however, we assume a zero inclination of the system and neglect the gravitational influence of 55~Cnc~B. Furthermore, we consider that the planets have zero obliquity. We consider the tidal forces between the star and the planets but we neglect the tidal interaction between planets. 
In this formalism, the total tidal force exerted by the star on the $i^{th}$ planet gives an acceleration for the planet of:
\begin{equation}\label{acc}
\begin{split}
\Mi \mathbf{a}_{i,{\rm tides}} &= \Bigg[ -\frac{3\G}{r_i^7}\left(\Mi^2k_{2,\star}\Rs^5+\Ms^2k_{2,i}\Ri^5\right) \\ 
& \quad\quad  - \frac{27}{2}\frac{\dot{r_i}}{r_i^8}\left(\Mi^2\sigma_{\star}\Rs^{10}+\Ms^2\sigma_i\Ri^{10}\right)\Bigg] \mathbf{e}_{r,i} \\
& \quad + \frac{9}{2r_i^7} \Big[\Mi^2\sigma_{\star}\Rs^{10}(\Os - \dot{\theta_i}) \\
& \quad\quad +  \Ms^2\sigma_i\Ri^{10} (\Omega_i - \dot{\theta_i})\Big]\mathbf{e}_{\theta,i},
\end{split}
\end{equation}

where $\Ri$ is the radius of the $i^{th}$ planet, $\Mi$ its mass, $\sigma_i$ its dissipation factor, $k_{2,i}$ its potential Love number of degree $2$. $r_i$ is the distance between the center of star and the $i^{th}$ planet and $\dot{r_i}$ its derivative. $\dot{\theta_i}$ is the derivative of the $i^{th}$ planet's true anomaly $\theta_i$ (the instantaneous orbital angular velocity). The radial vector $\mathbf{e}_{r,i}$ links the center of the star S to the center of the $i^{th}$ planet ${\rm P}_i$,  $\mathbf{e}_{r,i} = \frac{\mathbf{SP}_i}{|\mathbf{SP}_i|}$.  The orthoradial vector $\mathbf{e}_{\theta,i}$ is such that $\mathbf{e}_{r,i}\times\mathbf{e}_{\theta,i} = \frac{\mathbf{h_i}}{|\mathbf{h_i}|}$, where $\mathbf{h_i} = \frac{\Ms\Mi}{\Ms+\Mi}\mathbf{r}_i\times \mathbf{v}_i$ is the orbital angular momentum of the $i^{th}$ planet, here in heliocentric coordinates.

Here, we consider that only the two inner planets feel the effects of tides. This assumption is justified by the order of magnitude calculation made in section \ref{damping}. Both planets are assumed to be in pseudo-synchronous rotation. 

We also add the orbital acceleration due to General Relativity \citep{Kidder1995,Mardling2002}. Adding this force causes a change in the period of the eccentricity oscillations but does not change much the equilibrium eccentricity of planet e.

We performed different simulations with two sets of initial conditions. For each set of initial conditions, we varied the planetary dissipation from $10^{-5} \times \sigma_p$ to $100 \times \sigma_p$.

The first set of initial conditions is from \citet{Dawson2010} and the second one is from \citet{Endl2012}. Table \ref{table2} and Table \ref{table3} show the initial conditions for  \citet{Dawson2010} and \citet{Endl2012} respectively.

\begin{table}[h]
\begin{center}
\begin{tabular}{ccccc}
\hline
\hline
Planet & $M$ sin $i$ & $a$  (AU)  & $e$ & $\omega$ (deg) \\
\hline
55~Cnc~e & $8.26~\Mearth$  &  $0.01560$ & $0.17$ & $181$ \\
55~Cnc~b & $0.825~\Mjup$   & $0.1148$    & $0.010$ & $139$ \\
55~Cnc~c & $0.171~\Mjup$  &  $0.2403$ & $0.005$ & $252$ \\
55~Cnc~f & $0.155~\Mjup$   & $0.781$    & $0.3$ & $180$ \\
55~Cnc~d & $3.82~\Mjup$    & $5.74$    & $0.014$ & $186$ \\
\hline
\end{tabular}
\caption{Minimum mass, and orbital elements of the three outer planets of the 55~Cnc system.The values of $M$ sin $i$, $a$, $e$ and $\omega$ come from table 10 of \citet{Dawson2010}.}
\label{table2}
\vspace{5pt}
\begin{tabular}{ccccc}
\hline
\hline
Planet & $M$ sin $i$ & $a$  (AU)  & $e$ & $\omega$ (deg) \\
\hline
55~Cnc~e & $8.37~\Mearth$  &  $0.01544$ & $0.0$ & $90$ \\
55~Cnc~b & $0.80~\Mjup$    & $0.1134$    & $0.004$ & $110$ \\
55~Cnc~c & $0.165~\Mjup$  &  $0.237$ & $0.07$ & $356$ \\
55~Cnc~f & $0.172~\Mjup$   & $0.77$    & $0.32$ & $139$ \\
55~Cnc~d & $3.53~\Mjup$    & $5.47$    & $0.02$ & $254$ \\
\hline
\end{tabular}
\caption{Minimum mass, and orbital elements of the five planets of the 55~Cnc system.The values of $M$ sin $i$, $a$, $e$ and $\omega$ are from Tables 2 and 3 of \citet{Endl2012}.}
\label{table3}
\end{center}
\end{table}

\subsubsection{Results}

\textbf{Orbits and masses from \citet{Dawson2010}}

\begin{figure}[h!]
	\begin{center}
	\includegraphics[width=9cm]{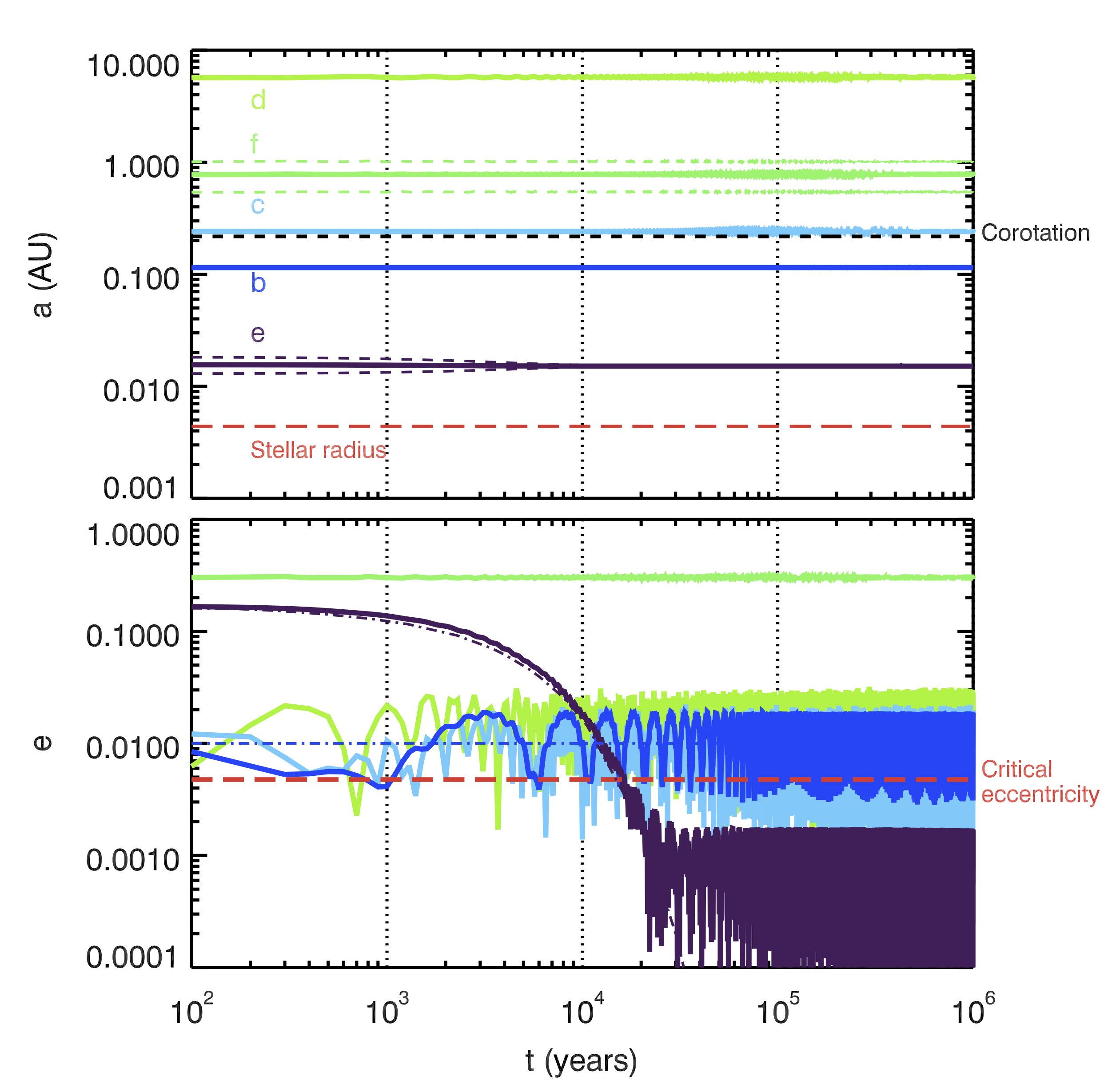}
	\caption{Evolution of semi-major axis and eccentricity for the five planets of 55~Cnc assuming the "nominal" tidal dissipation for 55~Cnc~e and the initial conditions come from \citet{Dawson2010}. Top panel: the semi-major axis is plotted in thick colored lines, the colored dashed lines represent the distance of aphelion and perihelion for each planet. The corotation radius is indicated by a black dashed line and the thick red dashed line is the stellar radius. Bottom panel: the colored full lines represent the eccentricities of the five planets, the dashed dotted lines represent the secular evolution  55~Cnc~e (purple) and 55~Cnc~b (blue) would experience if alone in the system. The thick red dashed line represent the critical eccentricity of planet e above which its secondary transit depth is increased by more than 10\% by tidal heating.}
	\label{17191_evolution}
	\end{center}
	\end{figure}

Figure \ref{17191_evolution} shows the 1 Myr evolution of the 5-planet system starting from the orbits from Table \ref{table2} and for a dissipation factor of $1\times\sigma_p$ for 55~Cnc~e. For the first few $10^4$~yrs, the eccentricity of 55~Cnc~e (initially set to 0.17) decreases as expected by the secular evolution calculations (dashed-dotted purple line), as if it were alone in the system, save a few oscillations. But after this initial decay, the eccentricity stops decreasing and oscillates around a few $10^{-3}$, due to the simultaneous influence of tidal damping and eccentricity pumping by the other planets.

Over $1$~Myr, the semi-major axis of 55~Cnc~e does not vary significantly. It decreases from $1.560\times 10^{-2}$~AU to $1.515\times 10^{-2}$~AU in the first few $10^4$~yrs while the eccentricity is still high and then decreases to reach $1.512\times 10^{-2}$~AU at $1$~Myr.
As expected from the order of magnitude timescale, the eccentricity of 55~Cnc~b does not decrease over the integration time. However due to the presence of  other planets exciting its orbit, 55~Cnc~b's eccentricity oscillates around $0.016$.

We tested if assuming pseudo-synchronization for 55~Cnc~e and 55~Cnc~b is correct when their eccentricities oscillate around their mean value. When 55~Cnc~e has the maximum eccentricity its pseudo-synchronization period is of $17.949016$~hr, and when it has the minimum eccentricity, it is of $17.950161$~hr. The difference between these two values is of $\lesssim 0.01$\%, which is negligible for dissipation. So the pseudo-synchronization remains a good assumption. For 55~Cnc~b tidal interaction is weak so it keeps its initial rotation period throughout the evolution. We chose here to assume pseudo-synchronization. 

To validate our version of the N-body code Mercury with tides, we verified it could reproduce the tidal evolution of single planet systems simulated with the code used in \citet{Bolmont2011} and \citet{Bolmont2012}, which solves the tidal secular equations for the semi-major axis, eccentricity (equation \ref{Hansene} of this paper) and rotation spin evolution of planet and star. We tested the General Relativity effects with \citet{Mardling2007}'s calculations for the secular evolution of two-planet systems. However as they use the constant phase lag model to compute the tidal force, we cannot compare their tidal evolution calculations with our simulations, which are based on the constant time lag model. The recent literature has debated the applicability of the constant time lag and constant phase lag models. The constant phase lag model has been shown to be mathematically inconsistent \citep{Efroimsky2013}. On the other hand, \citet{Greenberg2009} showed that the response of a viscoelastic fluid reduces exactly to the constant time lag approach in the limit where the forcing frequency is smaller than the natural frequency of the system and where damping is weak (see his equations 8b and c in these limits). A real planet's response is certainly much more complex than the one of a viscoelastic fluid, meaning that the constant time lag approach must also be used with care. The constant time lag model is a linear theory but taking into account nonlinear terms can result in noticeable changes \citep{Weinberg2012}. However, we choose this model because we believe that it stands on more robust mathematical grounds than the constant phase lag model.
%


As the dissipation factor of planets is not well known, we decided to cover a wide dissipation factor range, from $10^{-5}\times \sigma_p$ to $100\times \sigma_p$. As seen in section~\ref{section1}, the dissipation factor affects the tidal heating and change the value of the eccentricity required to affect the depth of the secondary eclipse. In addition, changing the dissipation in 55~Cnc~e has two effects on its orbital evolution.
First, it has an effect on how fast its eccentricity decreases. For a dissipation factor of $0.1 \times \sigma_p$, the eccentricity reaches the ``equilibrium'' in $\sim 3\times10^5$~yrs, while for a nominal dissipation factor, it does so in $\sim 3\times10^4$~yrs, and $\sim 3\times10^3$~yrs for $10 \times \sigma_p$. This is in agreement with the order of magnitude calculation of section \ref{damping}. In fact, the eccentricity damping timescale $\tau_{ecc}$ scales as $\Tp$ so as $\sigma_p^{-1} \propto (k_{2,p}\Delta T_p)^{-1}$, so if the dissipation factor is twice the fiducial value $\tau_{ecc}$ is half the value calculated for the nominal dissipation factor. 

Thus for dissipations below $10^{-2} \times \sigma_p$ the ``equilibrium'' eccentricity is not reached in less than $1$~Myr because the system is still in the circularization period. For dissipation lower than $10^{-2}\times \sigma_p$, the minimum, average and maximum values of the eccentricity of Table \ref{table4} are calculated for the whole evolution. For dissipation bigger than $10^{-1}  \times\sigma_p$, the ``equilibrium'' eccentricity is reached in less than $1$~Myr. The minimum, average and maximum values of the eccentricity are then calculated when the eccentricity has reached the ``equilibrium''.

\begin{figure}[h!]
	\begin{center}
	\includegraphics[width=9cm]{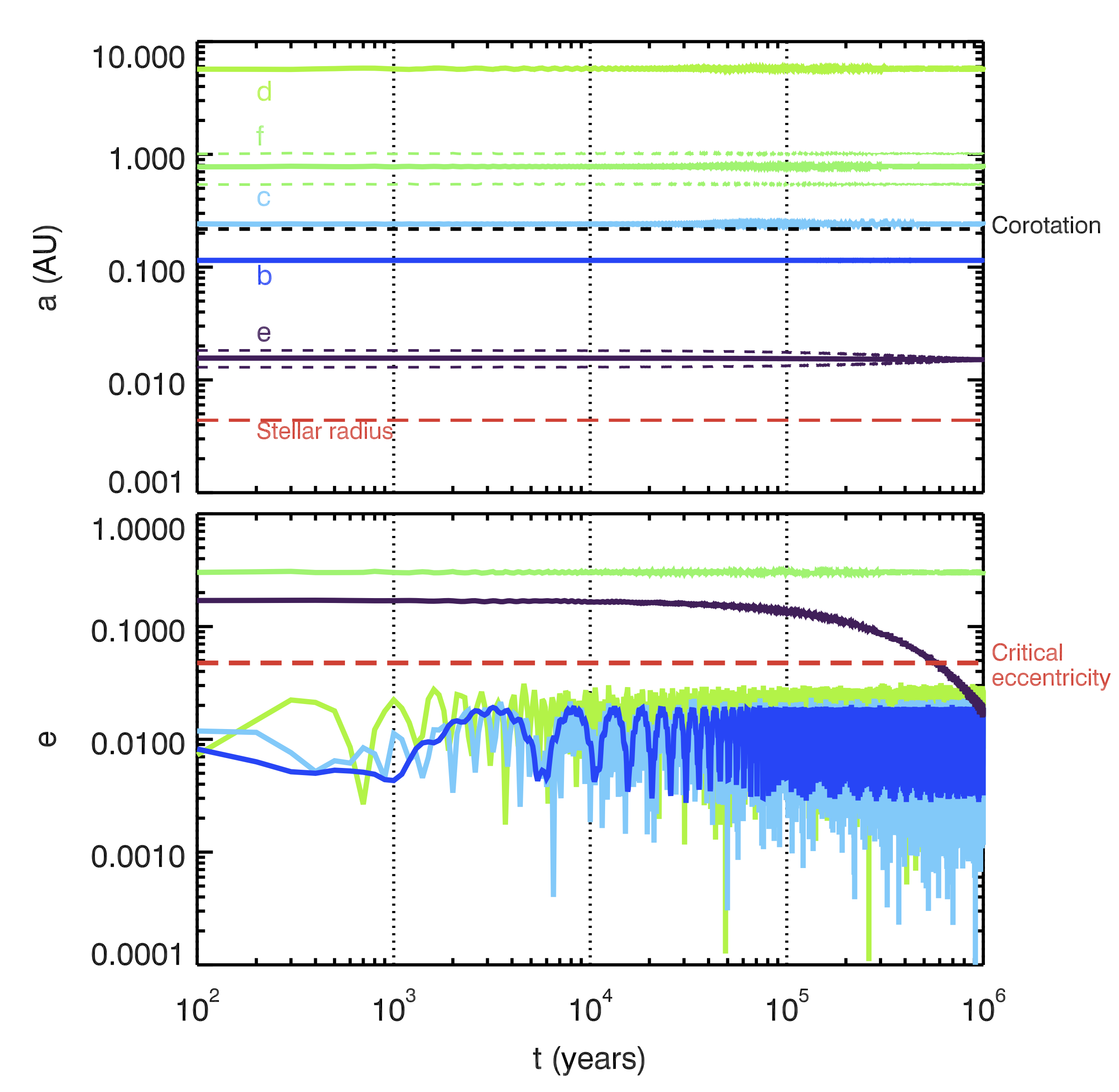}
	\caption{As Fig. \ref{17191_evolution} but assuming a dissipation factor of $10^{-2}\times\sigma_p$ for 55~Cnc~e.}
	\label{17192_evolution}
	\end{center}
	\end{figure}

The evolution of the system assuming 55~Cnc~e has a dissipation factor of $10^{-2} \times \sigma_p$ can be seen in Fig. \ref{17192_evolution}. The eccentricity of 55~Cnc~e remains higher than the critical eccentricity for $\sim 6\times10^5$~yrs. For a dissipation of $100\times \sigma_p$, the planet 55~Cnc~e falls on its star in $\sim1.2\times10^5$~yrs. It is therefore likely that the planet dissipates less than this value because the probability of observing the system in this configuration is low.

Changing the dissipation factor of 55~Cnc~e has also an effect on the forcing of its eccentricity by the other planets. The ``equilibrium'' eccentricity oscillates between a minimum and a maximum value, which are given in Table \ref{table4} for the different planetary dissipation factors. 

As expected, when the planetary dissipation is higher the forced eccentricity is lower. In Fig.~\ref{17191_evolution}, we plot the critical eccentricity above which the tidal heating has an effect of more than $10\%$ on the depth of the secondary transit. This value evolves with the dissipation. If the planet has a large dissipation factor, the eccentricity needed to have a strong tidal dissipation is lower. In section \ref{sumary}, we discuss the configurations that produce the strongest tidal signature in the transit depth.

\begin{center}
\begin{table*}[ht]
{\small
\hfill{}
\begin{tabular}{c|cccccccc}
&$10^{-5} \sigma_p$&  $10^{-4} \times \sigma_p$  & $10^{-3} \times \sigma_p$ &$10^{-2} \times \sigma_p$&  $10^{-1} \times \sigma_p$& $10^0 \times \sigma_p$ &$10^1 \times \sigma_p$ & $10^2 \times \sigma_p$ \\
\hline
\hline
min  ecc   & $1.66 \times 10^{-1}$ &$1.63 \times 10^{-1}$ &$1.33 \times 10^{-1}$& $1.56 \times 10^{-1}$ &$7.70 \times 10^{-5}$ &$5.00 \times 10^{-6}$& $6.70 \times 10^{-5}$& $1.00 \times 10^{-6}$\\
mean ecc& $1.70 \times 10^{-1}$&   $1.68\times 10^{-1}$ & $\mathbf{1.52 \times 10^{-1}}$& $\mathbf{6.82 \times 10^{-2}}$ &$1.75 \times 10^{-3}$&$9.36 \times 10^{-4}$&$3.16 \times 10^{-4}$&$4.45 \times 10^{-5}$\\
max  ecc  & $1.73 \times 10^{-1}$&  $1.73 \times 10^{-1}$&  $\mathbf{1.72 \times 10^{-1}}$& $\mathbf{1.71\times 10^{-1}}$ &$3.27 \times 10^{-3}$ &$1.80 \times 10^{-3}$&$6.85 \times 10^{-4}$&$9.10\times 10^{-5}$\\
crit. ecc    & $4.21 \times 10^{-1}$ & $2.71 \times 10^{-1}$ & $\mathbf{1.31 \times 10^{-1}}$ & $\mathbf{4.76 \times 10^{-2}}$&$1.52 \times 10^{-2}$ &$4.82 \times 10^{-3}$&$1.52 \times 10^{-3}$&$4.82 \times 10^{-4}$
\end{tabular}}
\hfill{}
\caption{Minimum, maximum and mean value of the eccentricity of 55~Cnc~e obtained in the simulation of the system proposed by \citet{Dawson2010} for different dissipation factor for 55~Cnc~e. The critical eccentricity needed to affect the planet is also indicated for each case.}
\label{table4}
\end{table*}
\end{center}

\begin{center}
\begin{table*}[ht]
{\small
\hfill{}
\begin{tabular}{c|cccccccc}
&$10^{-5} \sigma_p$&  $10^{-4} \times \sigma_p$  & $10^{-3} \times \sigma_p$ &$10^{-2} \times \sigma_p$&  $10^{-1} \times \sigma_p$& $10^0 \times \sigma_p$ &$10^1 \times \sigma_p$ & $10^2 \times \sigma_p$ \\
\hline
\hline
min  ecc  & $1.07 \times 10^{-4}$ &$1.22 \times 10^{-4}$ & $1.05 \times 10^{-4}$& $3.10 \times 10^{-5}$ &$2.50 \times 10^{-5}$ &$7.16 \times 10^{-4}$& $6.00 \times 10^{-6}$& $0.00$\\
mean ecc& $7.55 \times 10^{-3}$& $5.78\times 10^{-3}$ &  $4.36 \times 10^{-3}$& $4.33 \times 10^{-3}$ &$3.27 \times 10^{-3}$&$\mathbf{4.98 \times 10^{-3}}$&$8.16 \times 10^{-4}$&$1.74 \times 10^{-5}$\\
max  ecc & $1.30 \times 10^{-2}$& $1.11 \times 10^{-2}$&  $1.03 \times 10^{-2}$& $1.01\times 10^{-2}$ &$9.38 \times 10^{-3}$ &$\mathbf{1.04 \times 10^{-2}}$&$\mathbf{1.52 \times 10^{-3}}$&$4.54 \times 10^{-4}$\\
crit. ecc    & $4.21 \times 10^{-1}$ & $2.71 \times 10^{-1}$ & $1.31 \times 10^{-1}$ & $4.76 \times 10^{-2}$&$1.52 \times 10^{-2}$ &$\mathbf{4.82 \times 10^{-3}}$&$\mathbf{1.52 \times 10^{-3}}$&$4.82 \times 10^{-4}$
\end{tabular}}
\hfill{}
\caption{Minimum, maximum and mean value of the eccentricity of 55~Cnc~e obtained in the simulation of the system proposed by \citet{Endl2012} for different dissipation factor for 55~Cnc~e. The critical eccentricity needed to affect the planet is also indicated for each case.}
\label{table4}
\end{table*}
\end{center}


	\begin{figure}[h!]
	\begin{center}
	\includegraphics[width=9cm]{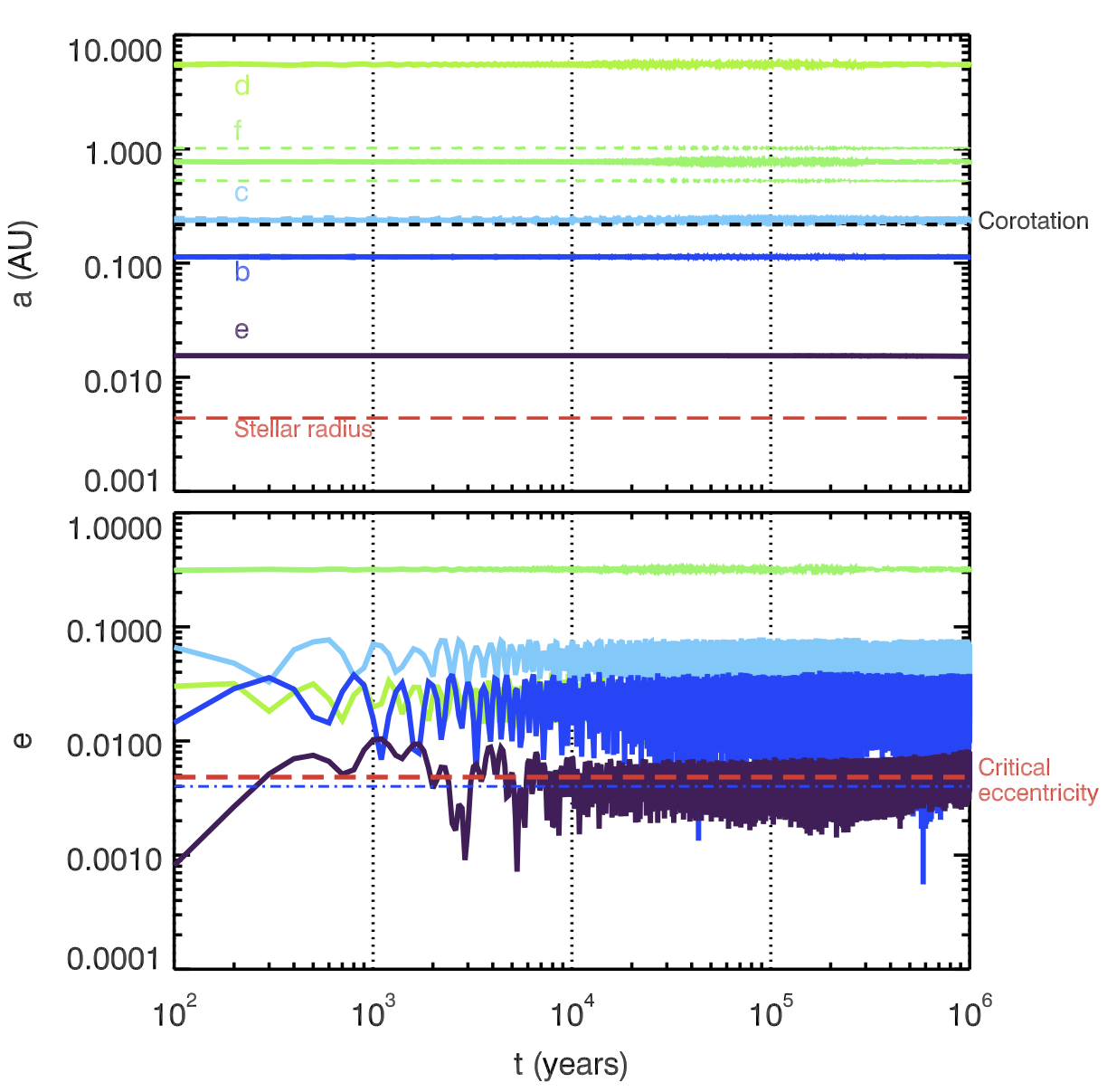}
	\caption{As Figure \ref{17191_evolution}, but the initial conditions come from \citet{Endl2012}.}
	\label{16341_evolution}
	\end{center}
	\end{figure}	

Fig.\ref{17191_evolution} shows that the eccentricity of 55~Cnc~e determined by \citet{Dawson2010}, $e=0.17$, is not stable as it is damped to values of a few $10^{-3}$ in less than $10^5$~yrs. Table~\ref{table4} also shows that the eccentricity of 55~Cnc~e never becomes high enough to cause a notable increase in the IR emission of the planet. In the absence of significant tidal heating, inefficient redistribution of heat and an albedo lower than 0.55 are required. 

\medskip

\textbf{Orbits and masses from \citet{Endl2012}}

Using the orbital distances, masses and exact arguments of pericenter proposed by \citet{Endl2012} lead to a destabilization of the system. We simulated the system with the N-body code Mercury without the tidal forces and without General Relativity and the outcome is the ejection of a planet. We also simulated the system without tides but with General relativity and the outcome is also the ejection of a planet. However, if we simulate the system with the orbital distances and masses from \citet{Endl2012} but using random arguments of pericenter (knowing that the values they provide are affected by significant uncertainties), we usually find that the system is stable on $10^7$~yrs timescales. 

Fig. \ref{16341_evolution} shows the simulation using orbital elements and masses from \citet{Endl2012}. The important differences between \citet{Endl2012} and \citet{Dawson2010} for the dynamical evolution of the system mainly lie in the values of the eccentricities of b and c. In \citet{Dawson2010}, b and c have small eccentricities: $0.01$ for b and $0.005$ for c, while in \citet{Endl2012} b has a small eccentricity: $0.004$ but c has a high eccentricity: $0.07$. In \citet{Dawson2010} after the damping of its initially high value, the eccentricity of e depends mainly on the eccentricity of b and in a weaker way on the eccentricity of c. However for \citet{Endl2012}, we observe that c excites b from the initial value of $0.004$ to the mean value of $\sim 0.02$. So e is excited by a more eccentric planet b than was the case for \citet{Dawson2010}, and in a weaker way by a more eccentric planet c.

The eccentricity of 55~Cnc~e cannot be zero because of the presence of the other planets, it oscillates around $4.98\times 10^{-3}$. The eccentricity of b also increases to oscillate around $2.20\times10^{-2}$. However, during the evolution of the five planets, the system frequently passes within the error bars of the current configuration derived by \citep{Endl2012}. For these configurations, the eccentricity of 55~Cnc~e is higher than the critical eccentricity for tidal heating. This simulation therefore provides dynamical evolution snapshots that are simultaneously consistent with the system properties derived from radial velocities, the secondary eclipse depth observed with Spitzer and an observable tidal heating.

As seen in Table~\ref{table4} and Fig. \ref{16341_evolution}, the mean eccentricity of 55~Cnc~e is also higher than the critical value. For $10 \times \sigma_p$, the maximum value of the eccentricity is above the critical eccentricity while for $0.1 \times \sigma_p$ the eccentricity always remain below the critical value.

In Fig~\ref{fig:A_e} (airless case) we can see that the upper limit on the albedo is $0.75$ for the mean eccentricity from this simulation. For the maximum eccentricity, this upper limit is as high as $0.85$. Interestingly, these high values of the albedo match those inferred for the exoplanet Kepler~10~b.

In the isothermal case (Fig~\ref{fig:A_e_unif}) the upper limit on the albedo is 0.1 for the mean eccentricity from this simulation. For the maximum eccentricity, this upper limit is 0.7.  No solutions are found for $0.1 \times \sigma_p$ and a marginal match is found with $10  \times \sigma_p$ with $A<0.05$.


\section{Tidal dissipation, transit depth and eccentricity}
\label{sumary}

\begin{figure}[h!]
	\begin{center}
	\includegraphics[width=9cm]{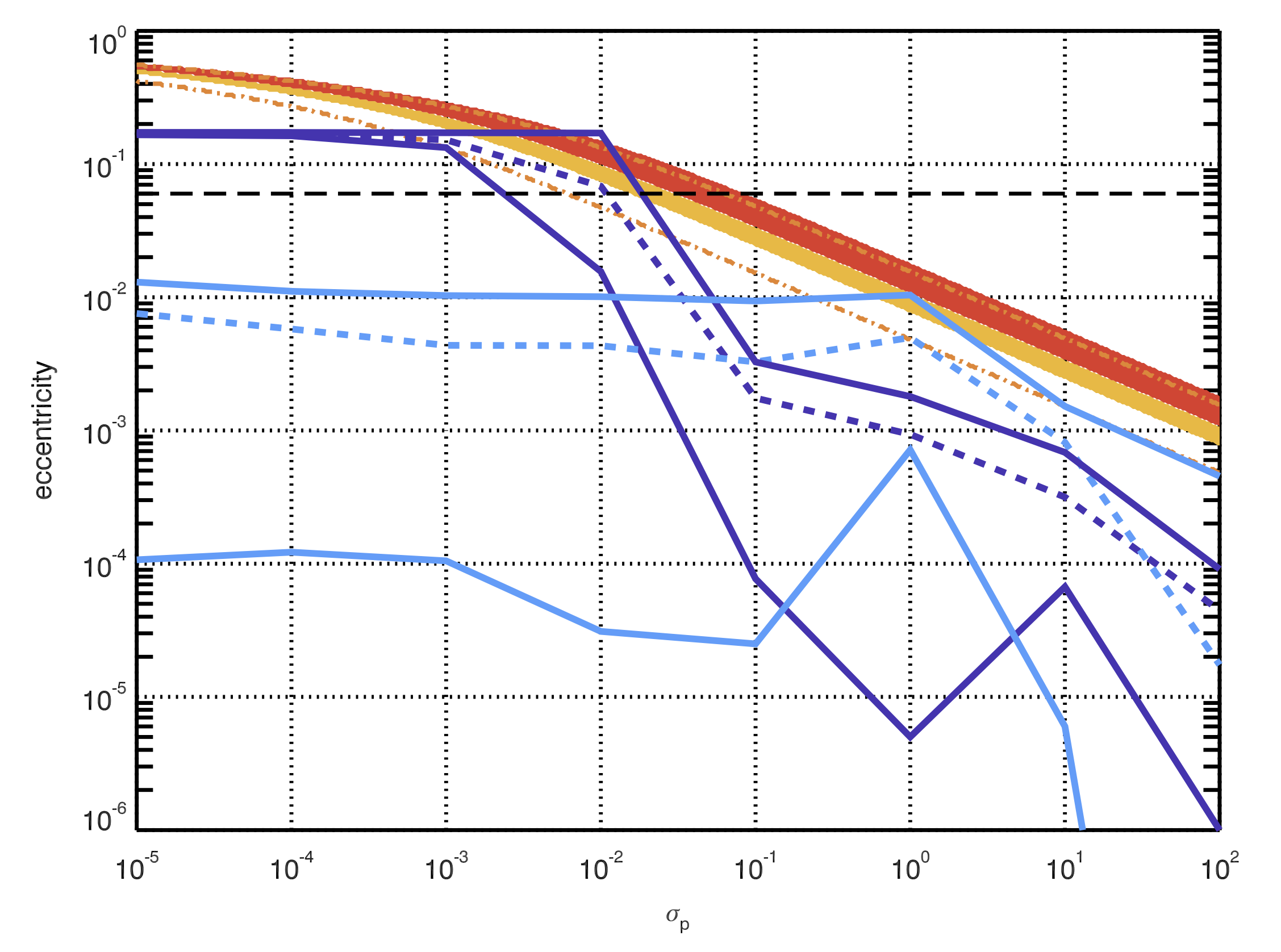}
	\caption{Eccentricity vs dissipation. The thick colored lines corresponds to the minimum, average and maximum eccentricity --  bottom full line, middle dashed line and top full line -- of 55~Cnc~e obtained with the simulations of \citet{Dawson2010} in purple, and \citet{Endl2012} in blue. The horizontal black dashed line corresponds to the upper limit from observations ($e=0.06$, D12). The color bands correspond to the region where the two conditions of equation \ref{twocond} are met: where the tides contribute to more than $10\%$ of the planet's flux and where the resulting transit depth is between the observational constraints. The light orange band corresponds to an albedo of $0$ and the red one to an albedo of $1$. The orange dashed dotted lines delimit the band corresponding to an albedo of $0.65$.}
	\label{map_sigma_ecc_better}
	\end{center}
	\end{figure}

Fig. \ref{map_sigma_ecc_better} sums up what has been discussed so far. It shows the range of simulated eccentricities of 55~Cnc~e for an evolution of $1$~Myr superimposed with the albedo-dependent region in the eccentricity-planetary dissipation parameters space corresponding to the two following conditions:

\begin{equation}\label{twocond}
\begin{cases}
 F_{{\rm p,tides}}/F_{{\rm p,tot}} > 0.1 \\
103~{\rm ppm} < F_{{\rm p,tot}}/F_\star < 159~{\rm ppm}
\end{cases}
\end{equation}

The first condition is that the tidal flux is more than $10\%$ the total flux of the planet. The second condition corresponds to the observability constrain of a transit depth of $131 \pm 28$~ppm, where $F_\star$ is the flux of the star calculated -- as the flux of the planet -- at a distance of $12.34$~pc (D12) in the Spitzer band.

The configurations for which the two conditions are met are therefore where the thick colored curves representing the output of our Mercury code intersect the shaded areas. For \citet{Dawson2010}, the overlap occurs from a dissipation of a few $\sim10^{-4}\times\sigma_p$ to a dissipation of a few $10^{-2}\times\sigma_p$. However, as the eccentricity of the planet has been constrained by observations (D12) to be smaller than $0.06$, the lower limit of the dissipation we can infer is about $8\times 10^{-3}~\sigma_p$ -- where the dashed dotted orange line intersects the horizontal black dashed line. For dissipations lower than $10^{-2}\times\sigma_p$, the eccentricity of 55~Cnc~e has not reached the ``equilibrium'' eccentricity, and the planet is still in its circularization period after $1$~Myr. The eccentricity decreases slowly with small oscillations as seen in Fig. \ref{17192_evolution}. So for dissipations lower than $10^{-2}\times\sigma_p$, the minimum eccentricity plotted in Fig. \ref{map_sigma_ecc_better} is the eccentricity of 55~Cnc~e at $t=1$~Myr and the maximum eccentricity is the high initial one $0.17$. It means that for these configurations, the two conditions of equation \ref{twocond} are met for a short period of time only. For example, for a dissipation of $10^{-2}\times\sigma_p$, the two conditions are met for $\sim 6\times10^5$~yrs after the proposed present configuration \citep{Dawson2010}. We do not know the previous state of the system, but it is likely that with such low dissipation the ancient eccentricity was significantly higher than its present value.

The most favorable configuration is for a planet with $A\sim0.65$ as can be seen in Fig. \ref{fig:A_e}. This albedo corresponds to the band delimited by orange dashed dotted lines in Fig. \ref{map_sigma_ecc_better}.
Fig. \ref{fig:A_e} shows that if the tides contribute to more than $10\%$ to the planet's flux, and the eccentricity of the planet is less than $0.06$ and the transit depth is in agreement with the observations, the albedo of the planet would be between $\sim 0.5$ and $0.7$.

For \citet{Endl2012}, a consistency between the two conditions and the eccentricity range obtained with Mercury simulations is found for dissipations around $1\times\sigma_p$. Depending on the dissipation, the two conditions are in agreement with the whole range of albedo. Contrary to the case of \citet{Dawson2010}, the eccentricity oscillates around the ``equilibrium'' eccentricity for all the dissipation range values. So over time the system will be periodically in configurations for which the two conditions are met. Over an oscillation of the eccentricity, the two conditions are met when the eccentricity is the higher. For example, as can be seen in Fig. \ref{map_sigma_ecc_better}, for a dissipation of $10\times\sigma_p$, the only configuration which would be in agreement with the two conditions is when 55~Cnc~e reaches the maximum value of the eccentricity oscillations, and it would mean that the planet has an albedo of $0.65$.



\section{Discussion}
\label{Discussion}

This work shows that the albedo cannot be constrained directly from the eclipse depth due to the influence of tidal heating on the radiative budget. For a given dissipation constant there is, therefore, a degeneracy between albedo, eccentricity and the assumed heat redistribution scenario. This degeneracy can partially be broken by using precise orbital photometry, as shown in fig.~\ref{fig:transit_sim}. A measure of the eclipse depth and/or the phase curve a different wavelengths would provide an additional constraint on the temperature map (as shown for instance by Maurin et al., \citeyear{Maurin2012}) and thus on the heat redistribution as well as on the relative contributions of tidal and radiative heating. A measure of the secondary eclipse and/or the phase curve at visible wavelengths that are not affected by thermal emission would provide a direct measurement of the albedo and would also help breaking this degeneracy. The ratio between the dayside emission of the planet and the star is however very small: $A \times 17.5$~ppm and should be measured over a broad wavelength range in order to infer the Bond albedo. The photometric survey of 55~Cnc with MOST \citep{Winn2011} reveals a modulation of the luminosity of the system (in the $350-700$~nm window), at the orbital period of the planet and whose phase is consistent with a phase modulation. However, the amplitude of this modulation, $\sim170$~ppm is nearly 10 times larger than what could be attributed to the planet. In addition the light curve does not exhibit a significant drop when the secondary eclipse occurs. This points to a modulation of the stellar luminosity itself. \citet{Winn2011} suggest that that part of the stellar hemisphere facing the planet could be slightly fainter than the rest of the star due to magnetic interaction between the planet and star. 
Such a stellar modulation would also make extremely challenging the extraction of the planetary phase modulation. How better is the situation in the infrared still has to be evaluated.\\

Our calculations assume that the internal heat flux is released uniformly over the planet. The existence of one or a few hot spots could significantly change the results. As an extreme case, if all the heat is released in the dark hemisphere, it has no effect on the dayside emission and thus a weak effect on the secondary eclipse depth. This would however have a strong signature on the orbital photometry.\\

In this study, we modeled the orbital photometric signature of planet e as if it was alone in the system. \citet{Kane_55Cnc_2011} showed that, at visible wavelengths, the modulation due to planet 'b' may not be negligible compared with that of planet e, depending of course on the respective albedos of planets e and 'b'. Characterizing the properties of planet e by orbital photometry thus requires the ability to distinguish between the two components, which is an additional challenge. At thermal wavelengths, the situation should be different because planet 'b' is a Jovian planet and atmospheric circulation tends to attenuate the day/night temperature contrast and thus the amplitude of thermal phase curve. To estimate this temperature contrast we can compare radiative and dynamical timescales. The radiative cooling timescale is given by $\tau_{R}  = \frac{c_{P}P} {\sigma g T^{3}}$ where $c_P$ is the specific heat of H$_2$, $P$ is the pressure at the considered level and $T$ is the bolometric brightness temperature. Assuming a superrotating atmosphere (as that of Venus or as predicted for strongly irradiated atmospheres (see for instance Showman et al., \citeyear{Showman2009} and Selsis et al., \citeyear{Selsis2011})), a relevant dynamical timescale is given by $\tau_{D}=R/v$ where R is the planetary radius and $v$ is the average of the zonal wind. Let us use $P=0.1$~bar (a typical photosphere for infrared radiation), T=700~K (the equilibrium temperature of the planet for $A=0$), $R=R_\textrm{Jup}$. Using 3D hydrocodes, \citet{Selsis2011} found $v \sim 200$m/s for the mean zonal wind speed for a 10 bar CO$_2$ atmosphere and an equilibrium temperature of 390~K, while \citet{Showman2009} found one order of magnitude faster winds for HD~189733b ($T_{eq}=1200$~K). Assuming $200<v<2000$~m/s yields $1 < \tau_R/\tau_D < 10$, which implies an efficient heat redistribution between the day and night hemisphere and thus a rather flat phase curve. The change of equilibrium temperature between periastron and apoastron is less than 5~K, a difference that should be completely averaged out by the inertia of the atmosphere. The contribution of planet 'b' to the photometric variations of the system should thus be negligible at thermal wavelengths. \\


Observations in several infrared bands will be necessary to establish the radiative budget of the planet and demonstrate the existence of an excess of luminosity due to tidal heating. In particular, in the presence of an atmosphere, the measured brightness temperature can correspond to the physical temperature at any atmospheric level. This temperature can locally exceed the highest temperature that can be found on an airless planet with a null albedo. This can happen in two cases. First, if there is a strong greenhouse warming and if the observed band corresponds to an atmospheric window that probes the surface of the planet. This situation can be illustrated by the thermal phase curves modeled by \citet{Selsis2011} for CO$_2$-rich atmospheres. In their Fig.~6, for instance, one can see that the 8.7~$\mu$m band (an atmospheric window) exhibits fluxes that are higher in the presence of an atmosphere. However, expected components of a planetary atmosphere, like CO$_2$, CO, SO$_2$, H$_2$S, CS$_2$, have strong absorption bands in the IRAC2 spectral window, making unlikely for this band to probe the deep and hot atmospheric layers heated by greenhouse effect. An ``excess'' of brightness temperatures can also be observed in an opaque band probing high stratospheric layers subjected to a strong thermal inversion. In the present case this would imply that the atmosphere is sufficiently opaque over the whole 4 to $5~\mu$m for the emission to come from very low pressure layers. Indeed, on the dayside of a synchronously rotating planet, the convective plume powered by the stellar irradiation is drafted to very high altitudes and an inversion appears possible only in the upper atmosphere. 

In a recent work, \citet{Maurin2012} proposed to use multi-wavelengths orbital photometry to derive the albedo, radius and inclination of rocky planets on circular non-transiting short-period orbits. The present study shows that applying this technique requires to insure that tidal heating is not affecting the radiative budget of the planet. Planets subjected to the influence of a massive and eccentric companion could have a small non-measurable eccentricity sufficient to produce a significant tidal heating. 

In our simulations, we assume the pseudo-synchronization of the planet spin. For rocky planets, the very existence of such state has been recently contested by \citet{Makarov2012}, who suggest that the only likely rotation configurations are spin-orbit resonances. Taking into account a more realistic rheology for the rocky planets, they show that the seemingly lowest-energy state of pseudo-synchronization is actually unstable. In the present study, the eccentricities remain small enough so that the difference between the 1:1 spin-orbit resonance and pseudo-synchronization has a negligible effect on dissipation, at least after a few $10^4$~yrs. The evolution during the first few $10^4$~yrs of the simulation of the \citet{Dawson2010} system would, however, be different. Using the constant time lag model, \citet{Rodriguez2012} showed that even though spin-orbit resonances are possible for large eccentricities, planets could be in pseudo-synchronization for eccentricities smaller than $0.1$. This study takes into account the oblateness of the planet in a most accurate way than in our model but does not incorporate a realistic rheology for the terrestrial planets such as in \citet{Makarov2012}.

The initial conditions of our simulations are the present values of semi-major axis and eccentricity of the planets. However, planet e must have formed farther out and migrated in due to tidal evolution. Tides are a dissipative process so it is delicate to integrate the evolution back in time, the solution diverges quickly to a huge semi-major axis and an eccentricity close to $1$. To investigate a possible formation distance from the star would require to explore a range of semi-major axis initial conditions to try reproduce present values. Our simulations are time demanding, so exploring a huge parameters space -- including the planetary dissipation-- in order to find a appropriate configuration would require a lot of time. So, we chose here not to treat past evolution, but investigate the present dynamics of the system.

In this paper, we show the influence of the eccentricity of the outer planets on the eccentricity of the inner planet. In the case of \citet{Endl2012}, the presence of the eccentric planet 'c' can increase the eccentricity of e from the supposed observational value of $0$ to $5\times10^{-3}$. The presence of the other planets also influences the eccentricity of the inner planet, so the values of the inner planet equilibrium eccentricity could change if more planets exist in the system as might be the case according to \citet{Raymond2008}.
 

\section{Conclusions}
\label{concl}

Our study shows that the orbit of exoplanet 55 Cancri~e cannot be fully circularized due to the influence of the other planets in the system. The forced eccentricity depends on the unknown dissipation constant and on the detailed set of orbital elements for the system. Using an Earth-like dissipation and orbital parameters from \citet{Endl2012}, we find an eccentricity that oscillate around a mean value $5\times10^{-3}$, up to $8.5\times10^{-3}$. The resulting tidal heating is sufficient to affect the thermal emission that has been observed at secondary eclipse with Spitzer/IRAC2 by \citet{Demory2012}. \\

With $e=0$ (and thus no internal heating) we find the same constraints on the planetary albedo as \citet{Demory2012}: $A < 0.55$ in the absence of heat redistribution and no match in the case of a uniform temperature. Including tidal heating the upper limit on the albedo becomes 0.9 and 0.7 without heat redistribution and in the uniform case, respectively. With these new constraints, both 55~Cnc~e and Kepler~10~b can have a high albedo ($A>0.5$). \\

Assuming an Earth-like dissipation allows us to use the Spitzer measurement to put an upper limit on the eccentricity: $e<0.015$, which is significantly lower than the observational constraint from D12 ($e<0.06$). The actual dissipation is unknown but future observations (in particular muti-wavelength orbital photometry) will have the ability to better constraint the contribution of tidal dissipation and thus the eccentricity. Alternatively, a more accurate measurement of the eccentricity and of the inclination of the system would allow us to constrain the dissipation. \\

This shows the importance tidal heating on the radiative budget of close-in exoplanets, in particular in compact systems that include eccentric planets. This case illustrates how the dynamics of a whole planetary system can influence the retrieval of the intrinsic properties of one particular planet (here, for instance, the surface albedo). \\


\begin{acknowledgements} 
FS acknowledges support from the European Research Council
(Starting Grant 209622: E$_3$ARTHs). EB and SNR thank the CNRS's PNP program. We thank the referee Rory Barnes for helpful comments that improved the paper. \end{acknowledgements}


\end{document}